\documentclass[acmsmall]{acmart}

\acmJournal{TORS}
\acmVolume{00}
\acmNumber{JA}
\acmArticle{00}
\acmMonth{05}
\acmYear{2026}
\acmDOI{10.1145/3814614}

\renewcommand\footnotetextcopyrightpermission[1]{} 
\AtBeginDocument{%
  }

\received{1 October 2025}
\received[revised]{17 February 2026}
\received[accepted]{22 April 2026}

\usepackage[utf8]{inputenc}
\usepackage{amsmath}
\usepackage{graphicx}
\usepackage{subcaption}
\usepackage{enumitem}
\usepackage{booktabs}
\usepackage{todonotes}
\presetkeys
  {todonotes}
  {inline,backgroundcolor=yellow}{}

\usepackage{xcolor}

\begin{document}

\title{Controlled Personalization in Legacy Media Online Services: A Case Study in News Recommendation}

\author{Marlene Holzleitner}
\orcid{0009-0005-0715-922X}
\affiliation{\institution{University of Klagenfurt}
\city{Klagenfurt}
\country{Austria}}
\email{marlene.holzleitner@aau.at}

\author{Stephan Leitner}
\orcid{0000-0001-6790-4651}
\affiliation{\institution{University of Klagenfurt}
\city{Klagenfurt}
\country{Austria}}
\email{stephan.leitner@aau.at}

\author{Hanna Lind Jorgensen}
\orcid{0009-0009-8382-7516}
\affiliation{\institution{Schibsted}
\city{Oslo}
\country{Norway}}
\email{hanna.lind.jorgensen@schibsted.com}

\author{Christoph Schmitz}
\orcid{0009-0008-3218-7080}
\affiliation{\institution{Schibsted}
\city{Oslo}
\country{Norway}}
\email{christoph.schmitz@schibsted.com}

\author{Jacob Welander}
\orcid{0009-0006-0651-6704}
\affiliation{\institution{Schibsted}
\city{Oslo}
\country{Norway}}
\email{jacob.welander@schibsted.com}

\author{Dietmar Jannach}
\orcid{0000-0002-4698-8507}
\affiliation{\institution{University of Klagenfurt}
\city{Klagenfurt}
\country{Austria}}
\affiliation{\institution{University of Bergen}
\city{Bergen}
\country{Norway}}
\email{dietmar.jannach@aau.at}

\begin{abstract}
Personalized news recommendations have become a standard feature of large news aggregation services, optimizing user engagement through automated content selection. In contrast, legacy news media often approach personalization cautiously, striving to balance technological innovation with core editorial values. As a result, online platforms of traditional news outlets typically combine editorially curated content with algorithmically selected articles---a strategy we term \emph{controlled personalization}. In this industry article, we evaluate the effectiveness of controlled personalization through an A/B test conducted on the website of a major Norwegian legacy news organization. Our findings indicate that even a modest level of personalization yields substantial benefits. Specifically, we observe that users exposed to personalized content demonstrate higher click-through-rates and reduced navigation effort, suggesting improved discovery of relevant content. Moreover, our analysis reveals that controlled personalization contributes to greater content diversity and catalog coverage and in addition reduces popularity bias. Overall, our results suggest that controlled personalization can successfully align user needs with editorial goals, offering a viable path for legacy media to adopt personalization technologies while upholding journalistic values.
 \end{abstract}



\maketitle
\renewcommand{\shortauthors}{M. Holzleitner et~al.}

\section{Introduction}\label{sec:introduction}
News has long been a key application area for recommender systems, with early developments dating back to the mid-1990s~\cite{Billsus1999APersonal,Kamba1995Krakatoa,Sakagami1997LearningPersonal}. News recommendation presents unique challenges that set it apart from other domains. For instance, the item catalog is highly dynamic, the lifespan of articles is typically short, and user interests are often influenced by various contextual factors such as the time of the day or the day of the week. These complexities, combined with the practical importance of the task, have spurred significant research in this area over the past decades~\cite{Wu2023PersonalizedNRSSurvey,KarimiIPM2018}.

However, much of the existing literature on news recommender systems focuses on their application within news aggregator platforms such as Google News or Yahoo! News~\cite{Das2007GoogleNEws,Li2010contextual,Lui2010Personalized}. In contrast, the adoption of news recommender systems by legacy news organizations---such as traditional newspapers and public service media providers---has been comparatively limited~\cite{hendrickx2021newsrecommendersystems}. Several factors contribute to this slower uptake. Notably, journalistic values and organizational missions play a central role in shaping decision-making within legacy media~\cite{Bauer2024WhereAre}. These values---such as promoting diverse viewpoints or fostering democratic engagement~\cite{Heitz26112022Benefits,Vermeulen2022ToNudge,Vrijenhoek2021recommenderswamission}---can sometimes be at odds with other objectives, including enhancing the online user experience, increasing reader engagement, or boosting customer retention~\cite{Vandenbroucke2024It}.

Integrating such values and tradeoffs into automated systems presents a significant challenge. As a result, it remains common for legacy media platforms to rely primarily on editorial judgment when deciding which content to feature prominently, rather than deferring to algorithmic systems~\cite{hendrickx2021newsrecommendersystems,Vandenbroucke2024It, Moller19052022Recommended, Gulla2021Recommendingnews}. Nevertheless, to capture some of the benefits of personalization, hybrid curation approaches are typically adopted~\cite{hendrickx2021newsrecommendersystems,klimashevskaja2024empowering}. In these approaches, editors retain control over core aspects of the digital offering, while other parts are personalized based on users' historical preferences. Moreover, it is not unusual that additional business rules are applied after a traditional recommendation algorithm generates a candidate set of items~\cite{Hauck2025BalancedPublicService}. These rules might, for example, filter out content that is too old, overly niche, or confined to a narrow set of sections (categories). In this work, we refer to such an approach as \emph{controlled personalization}.

Overall, although the literature on news recommendation is extensive, it has predominantly focused on the use case of news aggregators. In contrast, relatively little is known about the effectiveness of controlled personalization, a strategy frequently employed by legacy media organizations. With this study, we seek to address that gap by presenting insights from an A/B test conducted at Aftenposten, a major Norwegian news provider. Over the course of one month, the test compared the organization's existing non-personalized ranking system with a variant in which 20\,\% of the ranking score was influenced by a personalized recommendation algorithm.

Although the degree of personalization was relatively modest, our findings show that even limited adaptation to users' past preferences can yield marked improvements across several key metrics. Notably, personalized recommendations made it easier for users to discover relevant content, as reflected by higher click-through-rates and reduced scrolling activity. An analysis of the per-article reading behavior showed that the recommendations not only raised the readers' interest, as indicated by clicking on the article, but that users actually also spent more time on reading them than in the non-personalized condition.

Beyond enhancing user experience, we also found that personalization influenced reading behavior in ways that support editorial objectives. For example, users engaged with a wider variety of unique articles, and their reading patterns showed reduced bias toward highly popular content compared to the non-personalized setup. In addition, the adopted personalization strategy leads to a more diversified exposure and consumption behavior of articles. The key implication of our findings is that controlled personalization offers a promising middle ground for legacy media organizations striving to balance editorial oversight with personalized content delivery.

The remainder of the article is structured as follows: In Section~\ref{sec:previous-works}, we review findings from prior A/B tests of news recommender systems reported in the literature. Section~\ref{sec:use-case} introduces the application use case and outlines the experimental setup. In Section~\ref{sec:results}, we present the detailed results and key insights from the A/B test. Finally, Section~\ref{sec:discussion} summarizes our findings and discusses organizational challenges in industry, while Section~\ref{sec:conclusion} concludes and offers directions for future work.

\section{Previous A/B Tests in News Recommendation}\label{sec:previous-works}
In Section~\ref{subsec:overview-prior-works}, we will first review a selection of prior A/B tests reported in the academic literature over the past two decades. Afterward, we highlight similarities and differences of these works compared to our present study in Section~\ref{subsec:relation-to-ours}.

\subsection{Previous A/B Test Reports}\label{subsec:overview-prior-works}
The large majority of the research published in the field of recommender systems relies on offline experiments, i.e., it is based on studies using historical data without any human evaluation. Reports on A/B tests using real systems are generally quite scarce. An overview of selected field studies in different application domains---with a focus on the business value of recommender systems---can be found in~\cite{jannachjugovactmis2019}. In the following, we will highlight a number of articles presenting A/B test outcomes in the \emph{news domain} which were published during the last two decades.

In an early work from 2007, \citet{Das2007GoogleNEws} discuss the inner workings of the personalization approach developed for the news aggregation service \emph{Google News}. Besides common challenges in the news domain, such as a highly dynamic item catalog, a particularity of Google News is the massive user base of the service as well as narrow response time constraints. To address these challenges, a corresponding scalable and content-agnostic collaborative filtering approach is presented, which combines clustering techniques and news article co-visitation statistics in a hybrid approach. The system was evaluated using live traffic on the site and compared to a non-personalized baseline method that recommends recently popular articles. Technically, \citet{Das2007GoogleNEws} did not perform an A/B test by running the models in parallel. Instead, recommendation lists were created by merging the top-n candidates from each model in an interlaced way. The experiment was run for several months, and \textbf{click-through-rates} (\textbf{CTR}) were analyzed afterward. On average, the hybrids led to CTR statistics that were 38\,\% higher than the popularity-based baseline. Only occasionally, for certain celebrity news, the popularity-based method led to higher CTR values.

The collaborative filtering method by~\citet{Das2007GoogleNEws} however, turned out to have certain limitations, in particular that it cannot quickly enough recommend new items for which no feedback is available. In addition, the method---due to a certain popularity bias---frequently recommended entertainment news also to users who never read such articles before. An alternative information filtering approach for the Google News service was thus explored by~\citet{Lui2010Personalized} in 2010. This approach is based (a) on a Bayesian model to predict individual user interests from their past activities and (b) the current news trends. For producing the final recommendations, the outcomes of this method are then combined with those of the collaborative technique from~\citet{Das2007GoogleNEws} in a hybrid approach. The hybrid was then benchmarked against the pure collaborative technique in an A/B test involving 10,000 users over a period of 34 days. As performance measures, the CTR on the recommendations, the CTR of the Google News homepage, and the frequency of visiting the web site were used. The experiment showed that the proposed hybrid model led to a 30\,\% increase in CTR for the recommended items compared to the pure collaborative method. Interestingly, as the overall CTR on the website did not increase, the improved recommendations apparently `stole' traffic from other, non-personalized parts of the service. Finally, it was found that with the new recommendation service, the website visiting frequency was increased by 14\,\%, which the authors interpret as a sign of increased user satisfaction.

Differently from the experiments that were made on a news aggregator website in~\cite{Das2007GoogleNEws,Lui2010Personalized}, \citet{Kirshenbaum2012ALive} studied the effects of personalized recommendations on an individual online magazine, \emph{Forbes.com}. Specifically, they benchmarked 20 models in parallel for a period of five months in 2011 and 2012. The models included baseline methods based on popularity, pure collaborative techniques, pure content-based ones, and several hybrids implementing several algorithmic variations. Ultimately, a hybrid technique that combines item-to-item collaborative filtering with term frequencies and a novel Bayesian adjustment approach led to the best results in terms of the CTR. Overall, this method led to a 37\% CTR improvement over a time-decayed popularity-based method, increasing the CTR value from about 1\% to 1.4\%. However, the authors mention that it turned out to be non-trivial to beat popularity-based methods, noting that the winning method not only incorporated external knowledge from Wikipedia, but also included a novel adjustment approach.\footnote{See related discussions of the effectiveness of popularity and recency based methods in~\cite{Said2014DoRecommendations,Ludmann2017Recommending}.} Besides the CTR, the authors also asked if users spend more time on the site when provided with good recommendations. While they do not report such time measurements in the article, they found that the sessions of users who adopt any of the recommendations comprise significantly more article clicks than sessions of users who did not.

Click-through-rates and session lengths in terms of viewed articles are also in the focus of the A/B tests by \citet{Garcin2014Offline} performed on the \emph{swissinfo.ch} news website in 2013/2014. Their A/B tests are different from the previously discussed ones and involve a \emph{session-based} recommendation method \cite{Jannach2021SessionRSHB} called \emph{Context Trees}, which is able to serve recommendations to anonymous users. Overall, the authors observed after two-weeks evaluation phases that their method could lead to CTR increases of up to 35\,\% over the baseline method, and to significantly longer visit lengths. The gains however depend on general visit lengths, and the proposed method is more effective the longer the sessions are. A particularity of Garcin et al.'s study is that they perform an offline/online comparison, finding that the offline evaluation significantly underestimates the performance of the proposed Context Trees method. Furthermore, they critically discuss the use of the CTR (alone) as a performance measure. Finally, they also highlight that the user interface design matters, reporting that the position of the recommendation widget on the website can substantially impact the observed CTR.

In their A/B test from 2017, \citet{terHoeveEtAl2017DoNews} enhanced the user interface of the Dutch news aggregation platform \emph{blendle.com} with explanations for the provided recommendations. Before the A/B test, the authors conducted a survey-based user study, which suggested that users prefer to have explanations, but have no particular preferences regarding what kind of explanation are shown to them. An A/B test was then run on the website for 24 days. During this period, different ``\textit{heuristically selected}'' justifications were shown to users in the treatment group. An analysis of the recorded user behavior however showed that adding explanations do not have any significant effect, i.e., users did not open articles more frequently in the treatment condition than in the control group. The authors hypothesize that one reason behind this lack of effect may lie in the visual presentation of the explanations, which were not displayed prominently enough.

The study by~\citet{Zheng2018DRN} from 2018 again entirely focused on algorithmic innovations. Specifically, unlike the other field studies discussed so far, does not rely on supervised \textbf{machine learning} (\textbf{ML}) models but on a deep \textbf{reinforcement learning} (\textbf{RL}) approach for news recommendations. The goal of the proposed method is to overcome common challenges such as (1) highly dynamic changes in the catalog of articles, (2) a too strong focus on current rewards, and (3) limited diversity of existing approaches. In their technical approach, besides click behavior, the authors also consider the users' general \emph{activeness} as a signal to be considered in the model. Both offline and online experiments were conducted in the context of a major commercial news recommendation application. The offline experiments indicated that the proposed model is favorable over more traditional supervised models. Similarly, an A/B test that was run for one month on the platform showed that the proposed RL method was both leading to higher CTR values and higher diversity of the read articles. Interestingly, not all individual technical innovations---including the consideration of user activeness and increased exploration---led to the expected results in terms of higher CTR or diversity. Only the combination of all techniques turned out to be slightly favorable over the ``vanilla'' deep RL network. RL techniques were later also explored in an A/B test by~\citet{Zhu2022Integrated}, who observed an 1.58\,\% CTR improvement over an existing complex rule-based policy. Multi-armed bandits, as a simplified form of RL, were used for example in the A/B test reported by~\citet{Misztal2019Trend}, where they led to marked improvements in engagement-related \textbf{key performance indicators} (\textbf{KPIs}). Generally, contextual bandits have been identified quite early as a promising approach to deal with the particular dynamics of news recommendation, see e.g.,~\cite{Li2010contextual,Li2011Unbiased} for works developed for the Yahoo! News website.

The A/B test conducted by~\citet{Lu2020BeyondOptimizing} in 2019 for a Dutch online newspaper in the financial economic domain particularly focused on journalistic values. Therefore, like~\cite{Zheng2018DRN}, this work goes beyond solely relying on click-through-rates for measuring the effectiveness of personalized recommendations. Technically, a content-based recommendation method was used both for offline experiments and an online test. The offline experiments were performed to gauge aspects like accuracy, diversity, and serendipity compared to editorially chosen articles. In the online test, the content-based algorithm was A/B tested against a variation that considered the journalistic value of \emph{dynamism} in the recommendation. Dynamism was mainly implemented by putting more focus on item recency in the recommendations. The A/B test was run for 10 days and involved about 1,108 readers. The analysis of the logged data showed that the treatment group received more recent articles as recommendations, thus leading to higher \emph{dynamism}. In addition, the recommendations in the treatment group turned out to be more diverse and serendipitous. At the same time, the accuracy of the recommendations has not significantly changed. Thus, the authors concluded that it was possible in their study to design recommendations that foster journalistic values without compromising their relevance for readers.

A more business-oriented question stood in the focus of the work of~\citet{Iizuka2021TheEffect}, published in 2021. They target a common user interface design of online news websites, where news articles are interleaved with ads. The authors hypothesized that the level of ad consumption depends on the quality of the surrounding news articles. A log analysis based on data obtained from the most popular news application in Japan showed that low-quality articles may often lead to a high CTR, but these articles are not read to the end, as is more often the case for high-quality articles. The categorization of articles into different quality categories was done with the help of experts and a set of heuristics. In the subsequent A/B test, the personalized recommendations by the existing system were restricted to high-quality articles in the treatment group. The ``million-scale'' A/B test then indeed showed that focusing on high-quality articles lead to a higher CTR, higher conversation rate and more sales. A deeper analysis revealed that the gains in these business metrics were not limited to certain subgroups, such as readers interested primarily in entertainment news. Furthermore, the effect was also not limited to certain types of ads, and an uplift was observed for all but two genres, dating apps and automobile.

In most recent years, the emergence of \textbf{Large Language Models} (\textbf{LLMs}) has had a major effect on recommender systems research~\cite{Wu2024ASurvey,Deldjoo2024AReview,huang2025surveyfoundationmodelpoweredrecommender}, including applications in the news domain~\cite{wang2025surveyllmbasednewsrecommender}. Directly using an LLM for generating recommendations at scale is often not feasible, e.g., due to latency constraints. Therefore, \citet{Xi2025Towards} propose a novel model-agnostic recommendation framework that is able to leverage the reasoning capabilities and the world knowledge encoded in LLMs while ensuring fast response times of traditional algorithms. In this framework, both item-related knowledge outside of the recommendation dataset and reasoning about preference patterns are extracted through prompts from the LLM. The resulting information is then encoded into a compact representation, which is finally combined with existing recommendation models. Besides an extensive offline evaluation, the resulting system was deployed in an A/B test on Huawei's news and music platforms. During the period of the A/B test, a 7\,\% increase in terms of the Recall metric was observed compared to the existing baseline. According to the authors, the proposed framework has been put into production after the test.

This concludes our discussion of selected A/B tests in the news domain reported in the literature over the years. We note that some other works exist that report insights from deploying news recommender systems to real users. \citet{Lim2022AiRS}, for example, use an A/B test to do an ablation study for their proposed method. \citet{Li2011Unbiased}, on the other hand, uses an online test to explore if the online results correlate with their findings obtained from their offline evaluation. A particular form of A/B testing was also implemented in the context of the CLEF NewsREEL challenge~\cite{Hopfgartner2016Benchmarking}. A `living lab' was implemented in this competition, where participants received incoming click data from real-world news platforms, which they could use to return suitable recommendations and which were then displayed on the news outlets. Interestingly, quite simple strategies based on article popularity and recency often turned out to be hard to beat in terms of CTR~\cite{Ludmann2017Recommending}. A more recent initiative to better support user research in recommender systems is the POPROX platform~\cite{Burke2024Poprox}. At the moment, POPROX supports personalized news recommendations---taken from a curated set of Associated Press articles---that are delivered through daily newsletters. The newsletter is distributed to readers who have consented to participate in research. As a result, even though the newsletter system is a real-world working application, experiments done with the platform have both features of a user study and an A/B test.

\subsection{Main Contributions Relative to Previous Works}\label{subsec:relation-to-ours}
Our brief review in the previous section shows that while results from A/B tests in industry are published from time to time, the literature on the topic remains scarce. The case study presented in this article aims at contributing further insights on the effects of news recommendation in practice.

Differently from several earlier studies, we do not focus on the situation of large news aggregators. Instead, we aim to better understand the effects of personalization on online platforms of legacy media companies like Aftenposten who face distinct challenges. One particularity for example lies in explicit consideration of journalistic values and editorial missions, leading to a need for controlled personalization. As such, our work shares similarities with the studies presented in~\cite{Kirshenbaum2012ALive,Lu2020BeyondOptimizing,Garcin2014Offline}, and aims at providing helpful insights for the many existing legacy media online services.

Compared to the study by~\citet{Lu2020BeyondOptimizing} who focused on \emph{dynamism} in the A/B test, our analyses are based on an experiment that involved many more users which was online for a longer period of time. Furthermore, we consider more than one metric to assess different quality dimensions of the recommendations. This also sets us apart from the studies presented in~\cite{Kirshenbaum2012ALive,Garcin2014Offline}, which---like most studies that were made on large platforms---primarily focus on click-through-rates or on ad-based revenue as in~\cite{Iizuka2021TheEffect}.

In particular for the studies that present A/B test results for large platforms, the main focus of the article is often on technical innovations, and the discussion of A/B test outcomes is frequently limited to one or a few paragraphs. In our work, in contrast, our aim is a detailed analysis of personalization impacts in various dimensions. Also, in almost all earlier works, it is often unclear if the successful system was deployed in production afterward. For our study, we can report that this was the case, which supports the trust in the A/B test outcomes reported in the article.

Finally, in terms of the user interface, in some previous studies~\cite{Kirshenbaum2012ALive,Garcin2014Offline}, recommendations were presented in an own widget as part of the website. The observed effects can be heavily influenced by the positioning of the widget~\cite{Garcin2014Offline} and the specifics of the end-user device (desktop or mobile). Furthermore, there can be side effects when the widget display content that is shown somewhere else already~\cite{Kirshenbaum2012ALive}. In our study, the recommendations are placed in the main feed of the online newspaper and are thus a core part of the user experience. To isolate effects of end-user devices, we concentrated our analyses on readers that used mobile devices.

\section{Use Case Specifics and A/B Test Settings}\label{sec:use-case}
Aftenposten is Norway's leading national newspaper, which provides an online service since the mid-1990s. Like many legacy media organizations, Aftenposten faces the challenge of designing an online newspaper that provides a personalized user experience and at the same time supports journalistic missions, such as promoting democracy or social responsibility, and business goals, like subscriber growth \cite{Vandenbroucke2024It, Bauer2024WhereAre, Lu2020BeyondOptimizing, KarimiIPM2018}. The organization's approach is to implement a controlled level of personalization, where certain parts of the online newspaper are manually curated by editors whereas other parts are algorithmically populated.

Section \ref{subsec:aftenpostenpersonalization} details Aftenposten's values, mission, online service and personalization approach. Afterward, we present the specifics of an A/B test designed to evaluate the impact of controlled personalization on paying readers---the core focus of this investigation---in Section \ref{subsec:abtestsettings}.

\subsection{Aftenposten's News Service and Personalization Strategy}\label{subsec:aftenpostenpersonalization}
\paragraph{History and Organizational Values}
Aftenposten, Norway’s largest national printed newspaper, was founded in 1860 and is owned by Schibsted, a Norwegian media group. With a heritage of over 165 years, the company exemplifies a traditional legacy media newspaper. The organization has also been operating an online newspaper since 1995. As of 2025, Aftenposten reaches 700,000 daily users, with a subscriber base of 175,000 online and 75,000 print subscribers. The media house’s stated maxim is to strengthen democracy and freedom of speech through trustworthy journalism.\footnote{\url{https://schibsted.com/our-brands/aftenposten/}, accessed September 15, 2025.}

\begin{figure}[ht]
 \hspace{-2.25cm}
\centering
\begin{subfigure}[b]{0.47\textwidth}
 \centering
 \includegraphics[height=6.5cm,keepaspectratio]{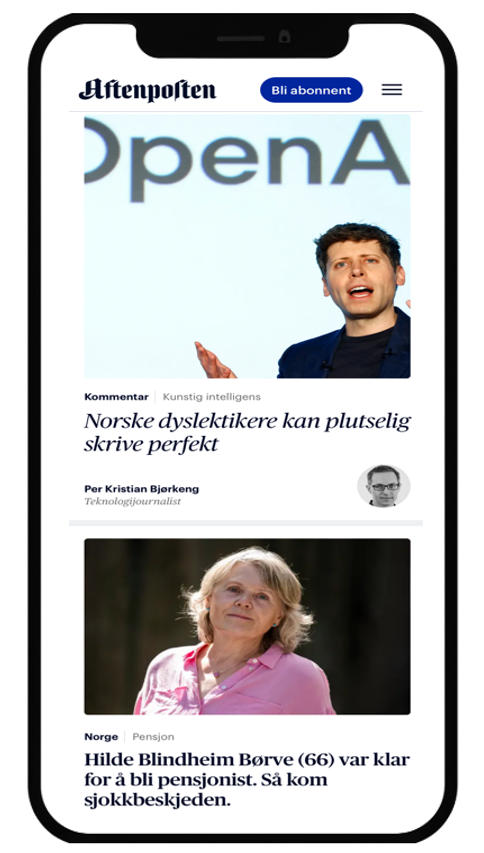}
 \caption{Front page of the mobile version.}
 \label{fig:subfig_front}
\end{subfigure}
\begin{subfigure}[b]{0.47\textwidth}
 \centering
 \includegraphics[height=6.5cm,keepaspectratio]{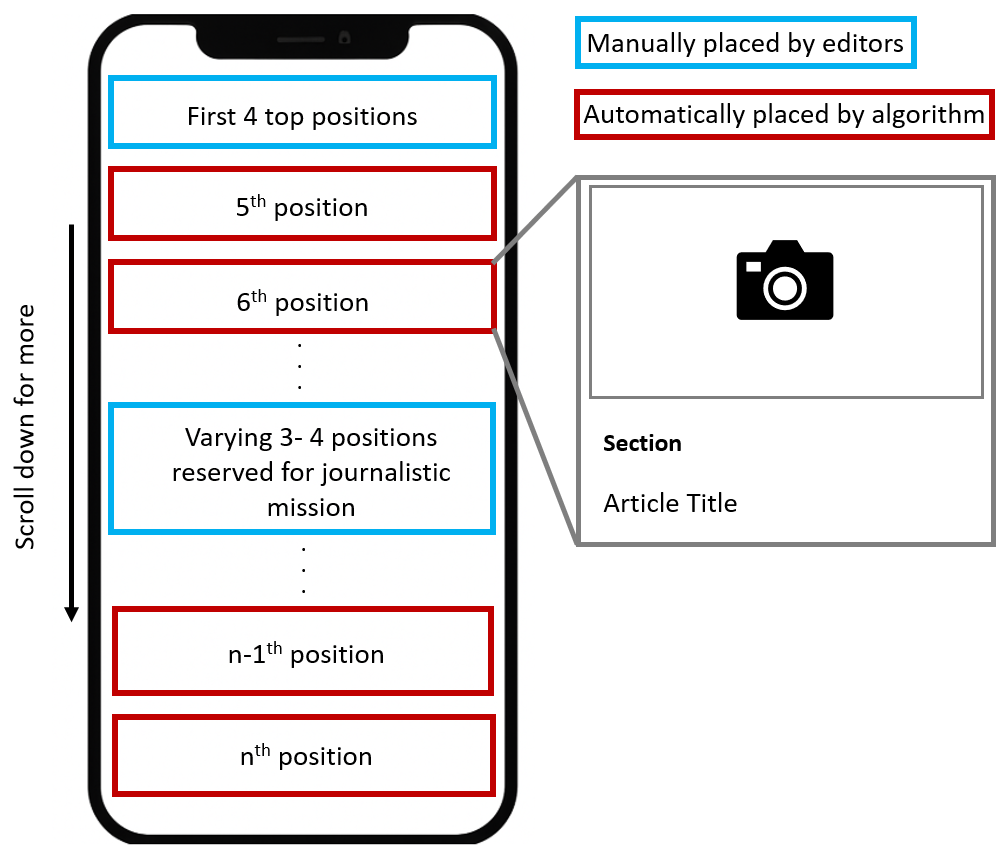}
 \caption{Stylized image of the user interface.}
 \label{fig:subfig_stylized}
\end{subfigure}
\caption{Structure of Aftenposten's mobile front page.}
\label{fig:mobile_and_stylized_version}
\Description[Structure of Aftenposten's mobile front page]{First, a screen capture of the mobile version is shown. Second, a stylized image illustrates the front page's structure: The top four positions and varying three to four additional positions are manually placed by editors. All other article positions are automatically placed by algorithms.}
\end{figure}

\paragraph{Aftenposten's Online Service and Front Page}
In this study, we focus on Aftenposten's online newspaper. It presents two distinct front page versions, tailored to desktop and mobile phone users. The user interface is designed around a continuous, scrolling stream of content. All articles and other content types are presented in a single feed---a long, dynamically updated list. Users scroll down within the news feed to access additional content. However, section navigation is also possible through the menu. Figure~\ref{fig:mobile_and_stylized_version}(a) shows a screen capture of the mobile version.

\paragraph{Article Placement}
The front page comprises areas that are manually curated by editors alongside those that are automatically populated. The editorial team manually positions the first four articles on the front page and, therefore, takes direct control over the content selection for the top positions. Furthermore, three to four additional articles on varying positions are manually curated to ensure the journalistic mission is met. While these positions are manually determined, the majority of the remaining front page articles---up to 90\,\%---are automatically selected. Prior to the experiments reported in this article, the algorithmic selection was always done in a \emph{non-personalized} way. This non-personalized selection is guided by a ranking algorithm that mainly considers an article's recency and popularity. An overview of the front page's structure, considering the article placement, is illustrated in Figure~\ref{fig:mobile_and_stylized_version}(b). The rules governing article placement are jointly defined by journalists and data scientists, but the editorial team retains the flexibility to override these rules.

\paragraph{Pool of Articles}
Editors exercise control over the algorithmic process by manually determining the pool of articles eligible for automated selection at a certain point in time. This manual process is guided by internal business rules, which ensures that a minimum number of articles of certain types, e.g., opinion articles, are considered in the pool. Other rules determine the lifetime of articles in certain sections.

\paragraph{Non-Personalized Ranking System}
In the existing non-personalized ranking approach, the resulting pool of articles is processed by three independent scoring algorithms, which rank the pre-selected content based on its popularity, recency, and past performance in terms of its CTR.
\begin{itemize}
\item The popularity algorithm ranks the eligible articles according to the number of clicks received within a defined time window, resulting in a scale from 0 to 100.
\item The recency algorithm ranks articles by their \emph{initial news value} in combination with their publication time, and outputs a score between 0 and 100. The initial news value is a number, set by the journalists when they publish it, describing the importance of the article. Overall, higher scores returned by this algorithm indicate greater initial value and/or a more recent time of publication.
\item The performance algorithm ranks articles based on their performance on the front page in terms of CTR over a given period, generating a score between 0 and 100.
\end{itemize}

The individual scores from each ranking algorithm are then weighted, summed, and normalized to a scale of 0 to 100, yielding a final composite score $cs$. More formally, let $\mathbf{s} = (s_1, s_2, s_3)$ be the individual scores for a given item, and $\mathbf{w} = (w_1, w_2, w_3)$ the corresponding weights, which determine the influence of each component on the overall ranking score for an article $i$. The overall score $\text{cs}_i$ of each article is computed as shown in Equation~\eqref{eq:non-personalized}\footnote{For readability, we omit the item index $i$ on the right-hand side of Equations~\ref{eq:non-personalized} and~\ref{eq:personalized}.}:
\begin{equation}
\label{eq:non-personalized}
{\text{cs}_i} = \frac{\sum_{n=1}^{3} w_n s_n}{\sum_{n=1}^{3} w_n}
\end{equation}

\paragraph{Personalization Strategy}
Starting in 2023, Aftenposten has been experimenting with personalization approaches that build on and extend the existing non-personalized ranking system. Online subscribers, i.e., paying readers, were selected as the target customer segment. The assumption is that for this subset of online users a sufficient amount of past activities has been recorded to support reliable personalization.

Technically, given a user $u$, the personalization approach consists of adding a fourth component to the existing scoring technique, as shown in Equation~\ref{eq:personalized}.
\begin{equation}
\label{eq:personalized}
{\text{cs}_{i,u}} = \frac{\sum_{n=1}^{3} (w_n s_n) + w_4 s_{4,u}}{\sum_{n=1}^{3} (w_n) + w_4}.
\end{equation}

The fourth component $s_{4,u}$ represents a score that reflects the assumed relevance of an article $i$ for a given reader $u$. The weight $w_4$ determines the influence of the personalized relevance score on the final composite score $\text{cs}_{i,u}$. Keeping the popularity and recency aspects in the equation, i.e., components one to three, is important, though, as these factors are central for effective news recommendations in practice~\cite{Ludmann2017Recommending}.

The personalized relevance score for a given user-item pair, i.e., the fourth component in Equation~\eqref{eq:personalized}, is based on a collaborative filtering model. Specifically, a matrix factorization technique~\cite{Koren2009Matrix} is applied on the user-article interaction matrix.\footnote{Details of the technique are omitted here for confidentiality reasons.} The model is retrained every three hours on filtered historical data from Aftenposten's front page, encompassing users' past behavior of the last 21 days. The filtering criteria are as follows: (i) only subscriber data is included,\footnote{Logged clicked data from anonymous users are thus ignored.} and (ii) articles must have accrued at least 100 clicks within the last 21 days. Following training, the collaborative filtering model generates a normalized ranking score for each user and article, scaled from 0 to~100.

\subsection{A/B Test Goals and Configuration}\label{subsec:abtestsettings}
\paragraph{A/B Test Goals}
The specific goal of the A/B test was to gauge the effects of personalized recommendations on business KPIs from two main categories: \textit{user engagement} and \textit{journalistic} values. The outcomes of the A/B test should then help deciding if the personalized recommendations should be permanently put in production. To reiterate, the primary objective of the case study reported in this article is to go beyond previous works like~\cite{Kirshenbaum2012ALive,Garcin2014Offline} that focus on click-through-rates as a primary and often only measure. Instead, we aim at considering also alternative quality and value-aware dimensions of news recommendations, as discussed in~\cite{Bauer2024WhereAre, Lu2020BeyondOptimizing}. In pursuit of this objective, before providing the detailed configuration of the A/B test, we first define the considered KPIs.

Table~\ref{tab:kpis} provides an overview of the KPIs that were considered in the two categories. KPI-1 and KPI-2 focus on traditional aspects of user engagement, i.e., how much of the exposed content\footnote{By exposure, we mean impression, i.e., users have seen the article because it was in the visible area of the screen at least.} readers did explore and how much content they were exposed to. These two KPIs are commonly seen as good proxies for business-oriented goals. Specifically, the assumption is that greater engagement leads to higher customer retention. For example, users who read more are considered less likely to cancel their subscription.

KPI-3 to KPI-7 instead reflect journalistic values, particularly the aims of broad information dissemination, promoting informed public discourse, and minimizing the formation of filter bubbles~\cite{pariser2011filter}. Specifically, the KPIs reflect what content is exposed to readers through the system and what readers actually consume. Three dimensions are considered in this context: content diversity, coverage, and article popularity. Higher diversity and coverage~\cite{Kaminskas2016Diversity} are assumed to better promote the described journalistic values. For article popularity, in contrast, lower values are considered better as they may help to avoid filter bubbles and feedback loops~\cite{Chen2023Bias, Klimashevskaia2024Survey}. Exact details of how the measurements were done will be provided later in the article.

\begin{table}[h!t]
\centering
\begin{tabular}{l p{0.2\linewidth}p{0.6\linewidth}}
\multicolumn{2}{l}{\emph{User engagement KPIs}} & \\
\midrule
\textbf{KPI-1} & Reader & Reflects subscriber loyalty and retention. \\
& Engagement & \\
\textbf{KPI-2} & User Impressions & Reflects scrolling and exploration behavior, needs contextual interpretation. \\ [0.8em]
\multicolumn{2}{l}{\emph{Journalistic value KPIs}} &\\ \hline
\textbf{KPI-3} & Topic Diversity & Reflects promotion of widespread access to information; \\
 & (Exposure) & instrumental in fostering an informed citizenry. \\
\textbf{KPI-4} & Topic Diversity & Reflects how well informed readers are, and represents the \\
 & (Consumption) & state of informedness in society. \\
\textbf{KPI-5} & Article Coverage & Reflects how broad the subscribers are informed. \\
 & (Consumption) & \\
\textbf{KPI-6} & Content Popularity (Exposure) & Reflects how often each article was shown to readers and helps discovering the emergence of feedback loops and popularity bias. \\
\textbf{KPI-7} & Content Popularity (Consumption) & Reflects the tendency of users to click on popular articles, helps identifying
feedback loops and popularity bias. \\
\bottomrule
\end{tabular}
\caption{Overview of Key Performance Indicators.}
\label{tab:kpis}
\end{table}

\paragraph{A/B Test Configuration}
The A/B test compared the existing non-personalized method (control) with the controlled personalized method (treatment). It was run from November 30th, 2023, to January 2nd, 2024, encompassing 34 days and approximately 58,000 subscribers. Paying readers were randomly assigned to the control or treatment variant, ensuring equal group sizes. The A/B test focused on the mobile platform due to its widespread use among the subscriber base.

\begin{table}[h!t]
\centering
\begin{tabular}{lccccc}
\hline
\textbf{Ranker} & $\mathbf{w_1}$ (Popularity) & $\mathbf{w_2}$ (Recency) & $\mathbf{w_3}$ (Performance) & $\mathbf{w_4}$ (Personal.) \\
\hline
Non-personalized & 0.50 & 0.25 & 0.25 & -- \\
Personalized     & 0.40 & 0.20 & 0.20 & 0.20 \\
\hline
\end{tabular}
\caption{Weight Distribution for Non-Personalized and Personalized Rankers}
\label{tab:ranker-weights}
\end{table}

The weights for the compared ranking models were set as shown in Table~\ref{tab:ranker-weights}. We note that also in the personalized condition, the main factors of the non-personalized ranker (popularity, recency, and recent performance) continue to play a major role, ensuring popular, fresh, and newsworthy content. The individual weights, and thus the relative influence of the components, were determined in a manual process based on the expertise of the team, consisting of a data analyst from the newspaper, a product manager, a deputy manager from the newsdesk, and the team working on Schibsted's content recommendation system Curate. Driven by the team of editorial, product, data, and experts from Curate, the implementation of news recommender systems at Aftenposten was a gradual and iterative process. They collaboratively selected ranking algorithm combinations, prioritizing a balance between personalization and serving the public interest. This was achieved through continuous monitoring of user responses, and crucially, regular review and validation of results with the newsroom, leveraging their years of experience with manual front page curation to inform implementation decisions. This approach moved beyond a purely algorithmic solution to incorporate editorial judgment and prevent solely relying on collaborative filtering. The cross-functional team decided to introduce personalized relevance scores that determine at most 20\,\% of the overall score, thereby implementing a conservative strategy. As the outcomes of the A/B test will however show, already incorporating a limited amount of personalization has a measurable effect on the resulting recommendations and on user behavior. Generally, putting too much emphasis on the personalization component has the risk that very recent or trending news might frequently be overlooked by readers in case they are outside their usual reading habits.

In terms of the data that was collected in the A/B test, we note that a fine-grained logging policy was implemented, which also sets this study apart from several previous works. Specifically, regarding user events, we not only recorded time-stamped click events on articles, which is the most central piece of information in prior works. Instead, we additionally recorded if an article was actually visible to users on the screen (counting as an impression), how far readers scrolled down in each article (giving us a reading percentage for each article) and how much time they spent on the article before moving to another one. Different types of article meta-data were collected as well, including the publication time of the articles, the article length, its actual content in terms of title and text, and the section (category) of the article.

\section{Impact Analysis}\label{sec:results}
In this section, we provide a detailed evaluation of the effects of personalization on user engagement and on journalistic values.

\subsection{Methodology}\label{subsec:methodology}
Before we discuss our results, we first review our data analysis methodology in terms of (i) data extraction and cleaning steps, (ii) the details of the used metrics, and (iii) the methods used for statistical significance testing.

\subsubsection{Data Preparation and Cleaning}
The data required for our analyses were extracted from Aftenposten's production database, covering the time span of the A/B test. The first and the last day of the A/B test were omitted as the test started and ended in the middle of the day. As mentioned earlier, the log data was limited to paying readers and to data that was generated through the mobile app. Importantly, only those paying readers were considered who were already established subscribers of the service.\footnote{A paying reader was considered established if he/she has been an active subscriber for at least $2$ months. Paying, but inactive readers, i.e., ones who did not click on a single article during the A/B test, were also excluded from the analysis.} This was done to ensure that sufficient historical information was available for the personalization approach already at the beginning of the A/B test. In terms of the considered articles, only events were taken into account for articles that were automatically selected and positioned, i.e., interactions with editorially curated content were excluded. We furthermore note that our data selection is restricted to articles that were clicked at least once during the A/B test, i.e., engaged content. As a result, the metrics that we use in our analysis do not consider impressions on articles that were never clicked.\footnote{For metrics such as CTR, including impressions from articles that were never clicked---that is, using the full impression log---would result in slightly lower absolute values than those reported in this article.} Finally, interactions with other content shown on the app, in particular videos, were not considered as well.

After extracting the data, a number of data cleaning steps were performed. Importantly, different heuristics were employed to identify and remove logged interactions that were likely not caused by human readers but by bots. The following bot detection patterns were applied.
\begin{itemize}
 \item \textit{High Event Rate}: Users with activity exceeding $13$ events per minute (above the $99.9$$^{\text{th}}$ percentile) were classified as bots and excluded from the dataset.
 \item \textit{Short Activity Duration}: Users exhibiting a high frequency of very short activity durations on clicked articles were removed. Specifically, we excluded users performing more than $9$ actions in less than $2$ seconds, using thresholds at the 5$^{\text{th}}$ and 99$^{\text{th}}$ percentiles, respectively.
 \item \textit{Stale Content Interaction:} Users clicking articles older than $10$ days more than $10$ times were removed. This was done knowing that the lifetime of most articles is very short, receiving the largest fractions of the interactions on the first day after publication.
\end{itemize}

Besides removing bot interactions, further consistency-preserving checks were applied. For example, we systematically searched for and removed incomplete data points, where the logging system had occasionally failed. Ultimately, the preprocessed dataset comprised a total of $1,\!621$ articles and over $56,\!000$ users, spanning approximately $2$ million interaction events (impressions and clicks). Table~\ref{tab:descriptive-statistics} presents the main descriptive statistics per experimental group, rounded to the nearest thousand.\footnote{Due to internal data protection regulations, we cannot share the raw or preprocessed datasets, and we only report rounded numbers. We acknowledge this limitation regarding reproducibility.}

\begin{table}[h!t]
\begin{tabular}{lp{3cm}p{3cm}p{4cm}}
\hline
& \textbf{Number of users} & \textbf{Number of clicks} & \textbf{Number of impressions} \\
 \hline
Control & $\sim 28$k   &  $\sim 336$k& $\sim 641$k \\
Personalization & $\sim 28$k   &  $\sim 339$k   & $\sim 563$k\\ \hline
\end{tabular}
\caption{Descriptive Statistics for the Experimental Groups.}
\label{tab:descriptive-statistics}
\end{table}

\subsubsection{Measurement Method}
This section provides details of the metrics and analyses that were used to assess the KPIs described above, separated into \emph{user engagement} metrics and \emph{journalistic value} analyses and metrics.

\paragraph{User Engagement Metrics}
These metrics are primarily based on the concept of \emph{impressions} and \emph{clicks}. An article \textit{impression} is recorded by the system when an article in the main feed was in the visible area of the user's screen. A \textit{click} means that the user tapped on the article in the main feed, which transferred the user to the actual article content. The following user engagement metrics were computed, which give us insights about the users' click and reading behavior. The metrics provide detailed assessments relevant to the user engagement KPIs listed in Table~\ref{tab:kpis} above (KPI-1 and KPI-2).
\begin{itemize}
\item \textit{Click-Through-Rate (CTR):} Indicates the proportion of seen articles that the user has clicked on. Higher CTR values indicate that the shown content raised interest by the user.
\item \textit{Canceled Click Rate (CCR):} Reports how many article views were `canceled'. A canceled click is defined as a click that has a \textit{reading percentage} (scroll depth) of less than $10$\,\% or an activity duration of $5$ seconds or less. High CCR values indicate that users find many articles irrelevant after clicking on them.
\item \textit{Impressions Per User (IPU):} Reports the average number of impressions per user during the testing period. Higher values indicate that the users explored more content, i.e., scrolled deeper in the feed.
\item \textit{Clicks Per User (CPU):} Reports the average number of clicks per user during the testing period. More clicks indicate more exploration of content.
\item \textit{Average Reading Percentage (RP) Per Click:} Indicates an estimate of how much of an article was read by a user, based on the scroll depth for an article~\cite{Iizuka2021TheEffect}. Higher percentages indicate that readers actually consumed more of an article's content.
\item \textit{Average Activity Duration (AD) Per Click:} Reports how long users were actively inspecting an article after clicking on it~\cite{xing2014beyond}. Longer durations may indicate higher user interest in the article.
\end{itemize}

Table~\ref{tab:metrics-user-engagement} summarizes how the \emph{user engagement} metrics are calculated.

\begin{table}[h!]
\centering
\begin{tabular}{rl}
  \toprule
   {Click-Through-Rate (CTR)} = & $\frac{\text{Number of clicks}}{\text{Number of impressions}}$ \\ [1em]
  {Canceled Click Rate (CCR)} = & $\frac{\text{Number of canceled clicks}}{\text{Number of clicks}}$ \\ [1em]
  {Impressions Per User (IPU)} = & $\frac{\text{Number of impressions}}{\text{Number of unique users}}$ \\ [1em]
  {Clicks Per User (CPU)} = & $\frac{\text{Number of clicks}}{\text{Number of unique users}}$ \\ [1em]
  {Average Reading Percentage Per Click} = & $\frac{\text{Sum of reading percentages of each click}}{\text{Number of clicks}}$ \\ [1em]
  {Average Activity Duration Per Click} = & $\frac{\text{Sum of activity durations of each click}}{\text{Number of clicks}}$ \\ [1em]
\bottomrule
\end{tabular}
\caption{Computation Rules for User Engagement Metrics.}
\label{tab:metrics-user-engagement}
\end{table}

\paragraph{Journalistic Value Metrics and Analyses}
The metrics and analyses in this area relate to three dimensions of interest: diversity, coverage, and popularity. The analyses are not only determined by the users' scrolling and clicking behavior, but often also by properties of the articles themselves. The metrics below directly correspond to the related journalistic value KPIs from Table~\ref{tab:kpis} above (KPI-3 to KPI-7).

\begin{itemize}
\item \textit{Diversity}
\begin{itemize}
    \item \textit{Distribution of Viewed and Clicked Items:} These analyses help us understand if the viewing and clicking behavior in terms of article sections is impacted by the recommendations~\cite{jannach2015what}.
    \item \textit{Gini Index:} The Gini index expresses the level of inequality in a distribution. Values close to 1 stand for extremely unbalanced distributions; a value of 0 expresses perfect equality and a uniform distribution. We use it to assess how the logged impressions and clicks are distributed across sections (article categories)~\cite{chong2019quantifiying,klimashevskaja2024empowering}.
\end{itemize}
\item \textit{Coverage}
\begin{itemize}
  \item \textit{Click Coverage:} This metric measures the fraction of impressed articles that receive at least one click. Higher values indicate that a larger fraction of the visible article pool has raised the reader's interest~\cite{hazrati2023choice, ge2010beyond}.
\end{itemize}
\item \textit{Popularity}
\begin{itemize}
  \item \textit{Average Recommendation Popularity (ARP):} This metric tells us how popular the recommended items are on average. Higher values indicate that the ranking system has a tendency to favor already popular items~\cite{yin2012challenging}.
  \item \textit{Average Click Popularity (ACP)}: The metric informs us about the average popularity of the recommended items that the user has actually clicked on. Higher values mean that readers more frequently click on already popular items~\cite{yin2012challenging}.
  \item \textit{Percentage of Popular Items (PPI)}: The metric reports the average percentage of popular items clicked per user. We used the $80/20$ Pareto rule to split the articles into popular and less popular items. Higher values indicate that users exhibit a tendency on focusing on more popular items~\cite{sonoda2022Analyzing}.
\end{itemize}
\end{itemize}

In Table~\ref{tab:metrics-journalistic-values}, we summarize how the \emph{journalistic value} metrics are calculated.

\begin{table}[h!]
\centering
\begin{tabular}{rl}
\toprule
   {Gini Index} = & \parbox{6.5cm}{$\frac{1}{2n^2 \bar{x}} \sum_{i=1}^{n} \sum_{j=1}^{n} \left| x_i - x_j \right| $ \newline {where $n$ is the number of sections, $\bar{x}$ is the mean of the number of impressions/clicks over all sections, and $x_i$ and $x_j$ are the number of impressions/clicks for sections $i$ and $j$}~\cite{chong2019quantifiying,klimashevskaja2024empowering}.}\vspace{1em} \\

  {Click Coverage} = & $\frac{\text{Number of unique clicks}}{\text{Number of unique impressions}} $ \\ [1em]
  {Average Recommendation Popularity (ARP)} = & $\frac{\text{Number of impressions}}{\text{Number of unique articles shown}}$ \\ [1em]
  {Average Click Popularity (ACP)} = & $\frac{\text{Number of clicks}}{\text{Number of unique articles clicked}}$ \\ [1em]
  {Percentage of Popular Items (PPI)} = & $\frac{\text{Sum of all clicked popular items per user}}{\text{Number of clicks}}$\\ [0.5em]
\bottomrule
\end{tabular}
\caption{Computation Rules for Journalistic Value Metrics.}
\label{tab:metrics-journalistic-values}
\end{table}

\subsubsection{Aggregation Strategies and Statistical Analysis}
We calculated aggregated values for all key metrics over the entire A/B testing period. For some analysis, we rely on daily statistics for a deeper understanding. Furthermore, for some analyses we categorized users according to their activity level~\cite{sonoda2022Analyzing}, to differentiate between highly active and less active users. The activity levels were determined by the number of clicks over the entire period. We selected the number of levels and the method by clustering users based on the optimization of the Calinski Harabasz Index~\cite{calisnki1974adendrite} and the Davis Bouldin Index~\cite{davies1979acluster}.

We applied different methods to test if any observed differences between the control and the treatment groups were statistically significant. Generally, we used a threshold of $\alpha=0.05$. Depending on the situation, we either used parametric or non-parametric tests. For metrics exhibiting approximately normal distributions (verified using Shapiro-Wilk tests and Q-Q plots for each variant) and homogeneity of variance (ratio of variances $<$ 2), we performed Student's \textit{t}-tests with Cohen’s \textit{d} as the effect size. A $95$\,\% \textbf{confidence interval} (\textbf{CI}) for the difference in population means was computed. When normality assumptions were met, but variances were unequal, we utilized Welch's \textit{t}-tests, again using Cohen’s \textit{d} as the effect size and $95$\,\% CI for the difference in population means. In case of violations of normality assumptions, Mann-Whitney U tests, with Cliff's delta as the effect size measure, were used. To quantify uncertainty, $95$\,\% bootstrap CI for Cliff's delta were computed. For some particular analyses, finally, we relied on Chi-squared tests (with either quantifying differences in proportions using a Wald $95$\,\% CI for $2x2$ contingency tables, or association strength using Cramér's V with $95$\,\% bootstrap CI) and permutation tests (computing the Jensen-Shannon divergence to assess differences in overall probability distributions, accompanied by $95$\,\% bootstrap CI). We report details of these particular tests when discussing the results of our analysis in the next section.

\subsection{Engagement Effects}\label{subsec:engagement}

\paragraph{Click Behavior}
Table~\ref{tab:click-behavior-statistics} shows the results for the \textbf{click-through-rate} (\textbf{CTR}), i.e., for the most common metric in the literature. We find that the CTR has markedly increased by over 14\,\% when the personalized ranking strategy was applied. A Chi-squared test of independence showed a significant association between experimental group and click behavior ($p < 0.01$, $95\,\% \,$CI for the difference in proportions$\,\left[-0.078,\; -0.075\right]$), although the effect size was small (Cramér's V $<0.08$). Meanwhile, the \textbf{canceled click rate} (\textbf{CCR}) remained constant. Combining these two observations indicates that the personalized article ranking not only led to more interest by users\footnote{We recall that a click on an article can only be interpreted as a sign of interest, but not as an indicator of the quality of the recommendation.} in the articles they have been presented, but also that the rate of articles that turned out to be irrelevant after the click did not increase. Let us note here that a clickbait strategy, where headlines and teasers are crafted to attract clicks, will often also lead to higher CTR values. However, with such a strategy, one might also expect higher cancel rates, when readers discover after clicking that the article was in fact not relevant or informative.

\begin{table}[h!t]
\centering
\begin{tabular}{lp{3cm}p{2cm}} 
\toprule
& \textbf{CTR} & \textbf{CCR}\\
& \tiny{Click-Through-Rate} & \tiny{Canceled Click Rate} \\ \hline
Control & 0.524 &   0.015\\
Personalization & 0.601**  (+14.60\,\%) &  0.015 ($\pm$ 0.0\,\%)  \\
\bottomrule
\end{tabular}
\caption{Click Behavior Statistics, ** means $p<0.01$.}
\label{tab:click-behavior-statistics}
\end{table}

\begin{figure}[ht]
\centering
\includegraphics[height=5.5cm,keepaspectratio, alt=]{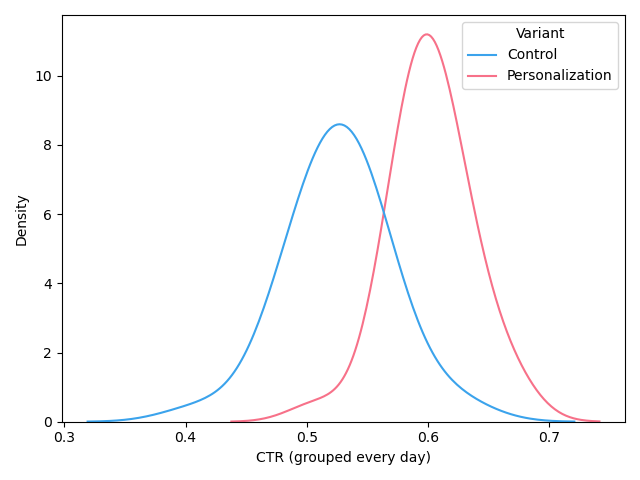}
\caption{Probability density function of daily CTR.}
\label{fig:pdf_ctr}
\Description[Probability density function of daily CTR]{The probability density function of daily CTR in the personalization group is shifted to the right, indicating higher CTRs than in the control group.}
\end{figure}

To obtain a deeper understanding of the effects of the personalization strategy on the CTR, we computed daily CTR statistics, compared the means, and plotted a \textbf{probability density function} (\textbf{PDFs}) as shown in Figure~\ref{fig:pdf_ctr}. The mean value of the daily CTRs increased from 0.525 ($SD=0.040$) to 0.601 ($SD=0.031$), representing a relative growth of $14.48\,\%$. A Welch's $t$-test revealed a statistically significant difference in daily CTR between the control and personalization group ($p<0.001$, two-sided $95\,\% \,\text{CI for the difference in population means}\,[0.058,\;0.094]$), with a large effect size (Cohen's $d$ $>2$), indicating relevant practical implications.

Diving deeper into the analysis, we compared impression and click statistics \emph{per user} in the control and treatment groups. The results are shown in Table~\ref{tab:ipu-and-cpu-metrics}. The analysis reveals that the increased CTR is due both to a \emph{decreased} number of impressions and a slightly \emph{increased} number of clicks per user. In particular, \textbf{impressions per user} (\textbf{IPU}) decrease by $11.70\,\%$, suggesting reduced user effort. Given the lower number of impressions, a decrease in \textbf{clicks per user} (\textbf{CPU}) might be expected. However, relative to the control group, CPU instead increases by $1.17\,\%$. We recall that a lower number of impressions means that users in the personalization group scrolled down less in the main feed of the application. The slight increase\footnote{In the context of online businesses, even small increases can be economically relevant. In ad-based businesses, for example, an increase of over $1\,\%$ directly translates into higher advertising revenue.} in CPU therefore suggests that, despite reduced scrolling, users were slightly more likely to find and click on interesting articles. A Mann-Whitney U test revealed that the observed differences for both metrics were statistically significant ($p<0.05$ and $p<0.01$ for IPU and CPU, respectively), although with a limited effect size (Cliff's delta $<-0.01$ and $<0.02$, one-sided $95\,\% \,$ bootstrap CI for Cliff's delta upper bound $=-0.0004$ and lower bound $=0.009$ for IPU and CPU, respectively). However, although the absolute effect sizes observed are modest, it is important to note that our experiment spanned only 34 days. Changes in user behavior resulting from personalization may require more time to fully manifest. Therefore, we consider it a highly promising outcome that we could observe significant and practically meaningful effects within this relatively short period.

\begin{table}[h!t]
\centering
\begin{tabular}{lp{3cm}p{3cm}} 
\toprule
 & \textbf{IPU}  & \textbf{CPU}   \\
 & \tiny{Impressions \mbox{Per User}} & \tiny{Clicks \mbox{Per User}}  \\ \hline
Control & 22.462&11.786   \\
Personalization & \textbf{19.834*} (-11.70\,\%)   &\textbf{11.924**}  (+1.17\,\%) \\
\bottomrule
\end{tabular}
\caption{Impression and Click Statistics per User, * indicates $p<0.05$, ** means $p<0.01$.}
\label{tab:ipu-and-cpu-metrics}
\end{table}

\paragraph{Reading Behavior}
Finally, we analyze how the personalized recommendation strategy impacts the users' reading behavior in terms of the \textbf{reading percentage} (\textbf{RP}) and the \textbf{activity duration} (\textbf{AD}) metrics. The results are shown in Table~\ref{tab:averaged-reading-behavior-metrics}. In either condition, about half of the articles was read after a click (RP). In the control group, readers on average consumed about 0.1\,\% more of the article content, i.e., a relative increase of 0.2\,\%. The time spent after a click (AD), in contrast, slightly increased in the treatment group, that is, 1.67\,\% in relative growth. For both metrics, the differences were statistically significant according to a Mann-Whitney U test ($p<0.05$ and $p<0.01$ for RP and AD, respectively), although with a very small effect size (Cliff's delta $<-0.003$ for RP and $<0.004$ for AD, two-sided $95$\,\% bootstrap CI for Cliff's delta $\left[-0.005,\; -0.00006\right]$ for RP and $\left[ 0.001,\; 0.006\right]$ for AD, i.e., the distributions for both experimental groups are almost similar in practice). Overall, combining the observations for the RP and AD metrics, there are no clear signs that the relevance or quality of the clicked articles is largely different in both experimental groups.\footnote{If we compute the daily averages for the RP and AD metrics, we find that the differences of the daily means are actually not statistically significant.}

\begin{table}[h!t]
\centering
\begin{tabular}{lp{3cm}p{3cm}} 
\toprule
 & \textbf{RP}  & \textbf{AD} \\
 & \tiny{Reading \mbox{Percentage}} & \tiny{Activity Duration (s)} \\ \hline
Control &   \textbf{48.941}* (+0.2\,\%)  & 103.363\\
Personalization &   48.825   & \textbf{105.085}**  (+1.67\,\%)  \\
\bottomrule
\end{tabular}
\caption{Reading Behavior Metrics, * indicates $p<0.05$, ** means $p<0.01$.}
\label{tab:averaged-reading-behavior-metrics}
\end{table}

\paragraph{Subgroup Analysis per Activity Level}
In the previous analysis we found that the average recorded activity level after a click is below two minutes. This is mainly due to a high number of users who exhibited limited interactions with articles.\footnote{A visualization of the distribution of activity durations of users can be found in the appendix in Figure~\ref{fig:ad_control} and in Figure~\ref{fig:ad_personalization}.} We iterate that per definition, scrolling times in the main feed are not counted in the AD metric. In order to understand if the observed impact on user engagement metrics can be equally found for subgroups of highly active and less active users, we conducted a subgroup analysis following ideas from \cite{sonoda2022Analyzing}.

\begin{figure}[ht]
\centering
\includegraphics[width=0.8\textwidth]{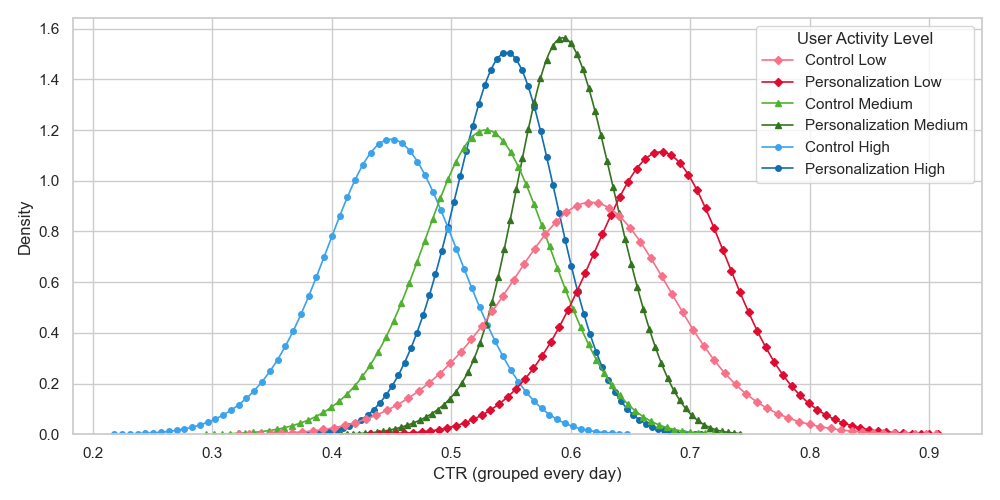}
\caption{Probability density functions daily CTR per user activity level.}
\label{fig:daily-ctr-user-activity-level}
\Description[Probability density functions daily CTR per user activity level]{The probability density functions of daily CTR per user activity level in the personalization group are shifted to the right, indicating higher CTRs than in the comparable segments of the control group.}
\end{figure}

Technically, we clustered the participants into three segments of low, medium and high activity as described earlier in Section~\ref{subsec:methodology} based on the number of clicks observed during the entire A/B testing period. We then computed daily CTR values for the segments to investigate if the personalized recommendations are effective for all types of users, see Table~\ref{tab:dailyCTR_IPU_user_activity_level}. A Welch's \textit{t}-test revealed that in every user activity segment, the daily CTR values were significantly higher in the personalization group ($p<0.01$, low: one-sided $95$\,\% CI for the difference in population means lower bound $=0.040$, medium: lower bound $=0.052$, high: lower bound $=0.079$) with strong effect sizes indicating practical significance (Cohen's $d$ $>1$). In relative numbers, the CTR of low, medium, and high activity users in the personalization group showed increases of over 9\,\%, 12\,\%, and 21\,\%, respectively, compared to the control group. Figure~\ref{fig:daily-ctr-user-activity-level} shows the probability density functions for this analysis.

\begin{table}[h!t]
\centering
\begin{tabular}{lp{3cm}p{3cm}} 
\toprule
 & \textbf{CTR}  & \textbf{IPU} \\
 & \tiny{Click-Through-Rate} & \tiny{Impressions Per User} \\
  \midrule
 Control Low & 0.611 & 1.617 \\
 Personalization Low & \textbf{0.672}** (+9.98\,\%) & \textbf{1.555}** (-3.83\,\%) \\
 \hline
 Control Medium & 0.527 & 2.775 \\
 Personalization Medium & \textbf{0.594}** (+12.71\,\%) & \textbf{2.556}** (-7.89\,\%) \\
 \hline
 Control High & 0.449 & 5.515\\
 Personalization High & \textbf{0.544}** (+21.16\,\%) & \textbf{4.481}** (-18.75\,\%) \\
  \bottomrule
\end{tabular}
\caption{Means of CTR and IPU per User Activity Level Measured Daily. ** indicates $p<0.01$.}
\label{tab:dailyCTR_IPU_user_activity_level}
\end{table}

Analogously, we investigated if a decreased number of impressions (IPU) can be observed across the user segments. As shown in Table~\ref{tab:dailyCTR_IPU_user_activity_level}, this was indeed the case. For example, we observed a decrease of over 18\,\% in impressions (IPU) in the high active user group compared to the control group. Again, a Welch's \textit{t}-test showed significant differences ($p<0.01$, low: one-sided $95$\,\% CI for the difference in population means upper bound $=-0.025$, medium: upper bound $=-0.120$, high: upper bound $=-0.776$) and strong effect sizes (Cohen's d $<-0.7$).

Finally, we analyzed if we find similar effects of an increased \textbf{activity duration} (\textbf{AD}) across the user segments of less active and higher active users. Interestingly, the analysis showed that a significant increase in activity durations can be observed both for low and high activity groups (a relative growth of 1.81\,\% and 2.71\,\%, respectively), with $p<0.001$ and $p<0.05$ according to Mann-Whitney U tests, although with small effect sizes (Cliff's delta $<0.01$; low: one-sided $95$\,\% CI for Cliff's delta lower bound $=0.002$ and high: lower bound $=0.0005$), indicating at least modest practical effects during the observation period. The activity durations of the medium group was however not significantly affected in the personalization group.\footnote{A visualization of the distribution of activity durations of users per activity level can be found in the appendix in Figure~\ref{fig:ad}.} Further investigations are needed in the future regarding this inconsistency of effects across user groups.

\paragraph{Summary}
Putting the click behavior and reading statistics together, we derive the following two main conclusions from our analyses:
\begin{itemize}
\item \textit{Personalization can help users identify relevant content more quickly}. Readers in the personalization group generate fewer impressions, i.e., they have to scroll down less to find relevant content, and the CTR is overall higher in the treatment group. The increased CTR values can be observed for users with different activity levels.
\item \textit{With personalized recommendations, users explore more articles}. This is evidenced by the increased average clicks per user in the treatment group.
\end{itemize}

Regarding the relevance of the clicked articles, no clear indications can be found that articles that were clicked on in the personalized condition were less relevant or informative than those that were clicked in the non-personalized condition. This rules out the hypothesis that the personalized recommendations have features of clickbait, where articles only raise attention but are not truly relevant. In terms of the reading behavior, no clear change can be observed. As was found in prior works \cite{Lui2010Personalized}, readers do not necessarily read more overall when receiving personalized recommendations.

Linking our findings to the user engagement KPIs in Table~\ref{tab:kpis}, we derive the following business-oriented implications. For \textit{KPI-1}, the personalization group shows higher reader engagement, reflecting stronger subscriber loyalty and retention. Despite the short 34-day observation period, positive effects are already evident, suggesting potential further gains over a longer horizon. Regarding \textit{KPI-2}, the personalization group experienced fewer user impressions, indicating reduced effort to discover interesting articles. This desirable outcome demonstrates that the personalized recommender effectively supports users in finding content with minimal effort.

\subsection{Journalistic Value Metrics and Analyses}\label{subsec:div-cov-pop}
We evaluated diversity, coverage, and popularity aspects to investigate the effects of personalization on the informedness of paying users, the breadth of the consumed content, and the potential of feedback loops and increasing popularity bias.

\begin{figure}[ht]
\centering
\includegraphics[height=7.6cm,keepaspectratio]{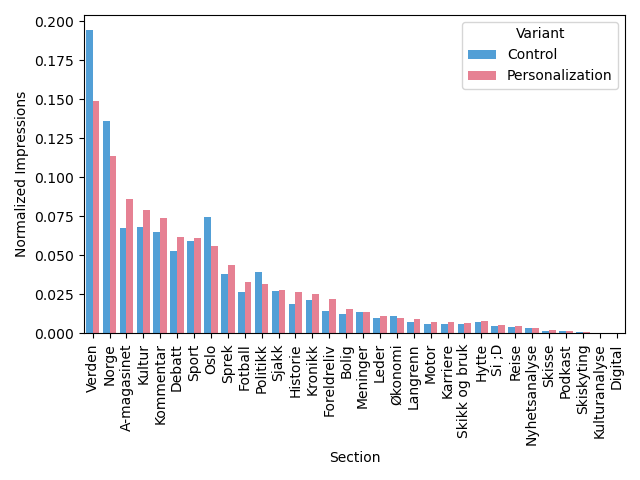}
\caption{Normalized distribution of impressions per section and variant.}
\label{fig:nimpressions}
\Description[Normalized distribution of impressions per section and variant]{The normalized impressions in the personalization group are slightly more evenly distributed across the spectrum of sections.}
\end{figure}

\paragraph{Diversity}
We analyze diversity aspects primarily by investigating how diverse the impressions and clicked articles are in terms of their sections (categories). Figure~\ref{fig:nimpressions} shows the normalized distribution plot of \textit{impressions} per section for the control group and the personalization group. A visual inspection suggests that the impressions (i.e., recommendations) in the personalization group are a bit more evenly distributed across the spectrum of sections. The distribution of the personalization group, for example has a less pronounced emphasis on the two most frequent (left-most) categories; also there seem to be more impressions on the less popular sections on the right-hand side.

A Chi-squared test was performed on the impression distributions of sections, resulting in a statistically significant difference ($p<0.001$), although with a limited effect size (Cramér's V $<0.11$, two-sided $95\,\% \,\text{bootstrap CI for Cramér's V}\,[0.101,\;0.105]$). The Gini index supports this finding of a statistical association between the experimental groups and the exposure of sections. A lower Gini index was identified in the personalization group (0.615 vs.~0.650), indicating a more equally distributed exposure of sections, corroborating a less concentrated and more diverse pool of articles. Given the higher total number of impressions in the control group, an additional permutation test was conducted to calculate the Jensen-Shannon divergence between the two probability distributions, i.e., the normalized impression distributions. A statistically significant difference between the two groups ($p<0.001$) was obtained, accompanied by a small effect size (Jensen-Shannon divergence $<0.06$, i.e., the two distributions are slightly different; two-sided $95\,\% \,\text{bootstrap CI for Jensen-Shannon divergence}\,[0.061,\;0.086]$). Overall, based on these analyses, we conclude that the personalization strategy leads to a more diversified exposure of articles. Again, we emphasize that even small effect sizes are encouraging, recognizing that stronger impacts of personalization on user behavior may only emerge over longer periods.

\begin{figure}[ht]
\centering
\includegraphics[height=8cm,keepaspectratio]{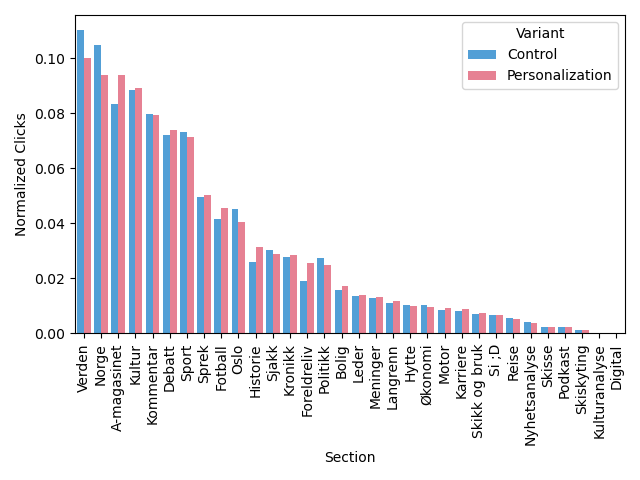}
\caption{Normalized distribution of clicks per section and variant.}
\label{fig:nclicks}
\Description[Normalized distribution of clicks per section and variant]{The normalized clicks in the personalization group are slightly more evenly distributed across the spectrum of sections.}
\end{figure}

Figure~\ref{fig:nclicks} shows the distribution of the \emph{clicks} per section. The findings for the clicks are in line with those obtained for the impressions. The click distribution is slightly more spread out across sections. The statistical test revealed a significant difference ($p<0.001$) between the distributions, although with a small effect size (Cramér's V $<0.05$, two-sided $95$\,\% bootstrap CI for Cramér's V\,$\left[0.044,\;0.048\right]$). The Gini index for the click data is again lower for the personalization group (0.542 vs.~0.552). A permutation test was again performed to calculate the Jensen-Shannon divergence on the normalized click distribution across sections, resulting in a significant difference ($p<0.001$) with a small effect size (Jensen-Shannon divergence $<0.03$, i.e., the two click distributions differ moderately; two-sided $95$\,\% bootstrap CI for Jensen-Shannon divergence $\left[0.024,\;0.03\right]$), confirming a slightly more diversified consumption behavior in personalization.

In practical terms, the personalized recommendations lead to the effect that the order of the most `popular' sections has changed, both in terms of impressions and clicks. Table~\ref{tab:norm_distr_sections} shows the top ten sections for the control and the personalization group. This parallel change suggests that (a) users frequently adopt the personalized suggestions and (b) that personalization can lead to a shift in consumption behavior.\footnote{Similar shifts in consumption patterns were observed in an early observational study in the e-commerce domain \cite{ZankerBricmanEtAl2006}.}

\begin{table}[ht]
\small
\centering
\begin{tabular}{lllll}
  \toprule
  & & \textbf{Sections} & & \\
  \hline
  & \textbf{Impressions} & & \textbf{Clicks} & \\
  \hline
Top 10 & Control & Personalization & Control & Personalization\\
  \midrule
 1 & Verden & Verden & Verden & Verden \\
 2 & Norge & Norge & Norge & Norge \\
 \textbf{3} & \textbf{Oslo} & \textbf{A-magasinet} & \textbf{Kultur} & \textbf{A-magasinet} \\
 4 & Kultur & Kultur & \textbf{A-magasinet} & \textbf{Kultur}\\
 \textbf{5} & \textbf{A-magasinet}   & \textbf{Kommentar} & Kommentar & Kommentar\\
 \textbf{6} & \textbf{Kommentar} & \textbf{Debatt} & \textbf{Sport} & \textbf{Debatt}\\
 7 & Sport & Sport & \textbf{Debatt} & \textbf{Sport} \\
 \textbf{8} & \textbf{Debatt} & \textbf{Oslo} & Sprek & Sprek \\
 \textbf{9} & \textbf{Politikk} & \textbf{Sprek} & \textbf{Oslo} & \textbf{Fotball} \\
 \textbf{10} & \textbf{Sprek} & \textbf{Fotball} & \textbf{Fotball} & \textbf{Oslo} \\
  \bottomrule
\end{tabular}
\caption{Most Popular Sections in Terms of Impressions and Clicks.}
\label{tab:norm_distr_sections}
\end{table}

\paragraph{Click Coverage}
Our click coverage analysis aims at assessing if personalization has an impact on the fraction of articles that were actually clicked on by users after having been displayed. We note that one limitation of our coverage analysis is that we do not dispose of knowledge about the entire available item catalog at a given point in time.\footnote{When determining a traditional coverage metric, we would compute the fraction of catalog items that are ever recommended relative to the full catalog.} Thus, we use impressions as a proxy for the available catalog. Our approximated catalog is therefore defined as the set of items that were ever displayed to users, that is, items appearing in the impression logs. Because this approximated catalog represents only a subset of the full catalog, the resulting click coverage values tend to be relatively high.

Generally, low click coverage can be the result when the personalized or non-personalized recommendations focus too much on a certain subset of the displayed pool of articles, e.g., only on highly popular ones. In different use cases, also in the media streaming domain, one purpose of a recommender system is thus to increase what Netflix calls the \emph{effective catalog size} \cite{Gomez2016Netflix}.

In Table~\ref{tab:ClickCoverageoverall}, we report the coverage of clicked articles over the entire A/B testing period. We again drill down to different customer segments of users with low, medium and high activity. Overall, we can observe that the coverage values are consistently very close to 1, which means that every single displayed article from the pool was clicked on at least once by some user in the A/B test period.\footnote{We note that traditional coverage metrics from the literature do not account for the frequency of item exposure or user interactions, and every single interaction may be sufficient to contribute to the coverage metric. This limitation also applies to our \textit{click coverage} metric, where it is highly likely that nearly every displayed article receives at least one click from one of the 28.000 users in each experimental group.} Overall, this suggests that users across all customer segments in both groups actively engaged with the displayed articles, rather than systematically ignoring visible content.

Clearly, the coverage problem for Aftenposten with its limited and constantly updated pool of articles is not as pressing as for companies like Netflix or Spotify, which have catalog sizes in the millions, with the additional problem that a few ``blockbuster'' items dominate the streaming statistics.

\begin{table}[!h]
\centering
\begin{tabular}{lp{3cm}}
  \toprule
 & \textbf{Click Coverage} \\
  \hline
 Control Low & 0.988 \\
 Personalization Low & 0.993 (+0.51\,\%) \\
 \hline
Control Medium & 0.993\\
 Personalization Medium & 0.995 (+0.20\,\%) \\
 \hline
Control High & 0.992\\
 Personalization High & 0.995 (+0.30\,\%) \\
\bottomrule
\end{tabular}
\caption{Click Coverage Measured per User Activity Level.}
\label{tab:ClickCoverageoverall}
\end{table}

While click coverage is very close to 1, when the data of the entire testing period is considered, the picture is slightly different when daily click coverage statistics are in the focus. Table~\ref{tab:ClickCoverage} shows the average of the \textit{daily} coverage statistics both for the control and the personalization group. Considering the daily click coverage in the personalization group, its variability is slightly lower than that of the control group (SD $= 0.023$ vs. SD $= 0.037$).\footnote{A visualization of the probability density functions of the daily click coverage per group can be found in the appendix in Figure~\ref{fig:pdf_click_coverage_kde}.} At this more detailed level of analysis, differences become visible. Specifically, we find that the daily click coverage in the personalization group is significantly higher than in the control group according to a Welch's \textit{t}-test ($p < 0.001$, one-sided $95$\,\% CI for the difference in population means lower bound $= 0.02$), with a large effect size (Cohen's $d$ $>1$). Specifically, we observed a relative increase of approximately 4\,\% compared to the control group. This finding indicates a more extensive or broader distribution of clicks across the content displayed in the personalization group, with high practical significance. Specifically, a larger fraction of the personalized visible article pool has raised the reader's interest.

\begin{table}[h!t]
\centering
\begin{tabular}{lp{3cm}}
  \toprule
 & \textbf{Click Coverage} \\
  \hline
 Control & 0.902\\
 Personalization & \textbf{0.938}*** (+3.99\,\%) \\
  \bottomrule
\end{tabular}
\caption{Means of Click Coverage Measured Daily. *** indicates $p < 0.001$.}
\label{tab:ClickCoverage}
\end{table}

\begin{figure}[ht]
\centering
\includegraphics[width=0.98\textwidth]{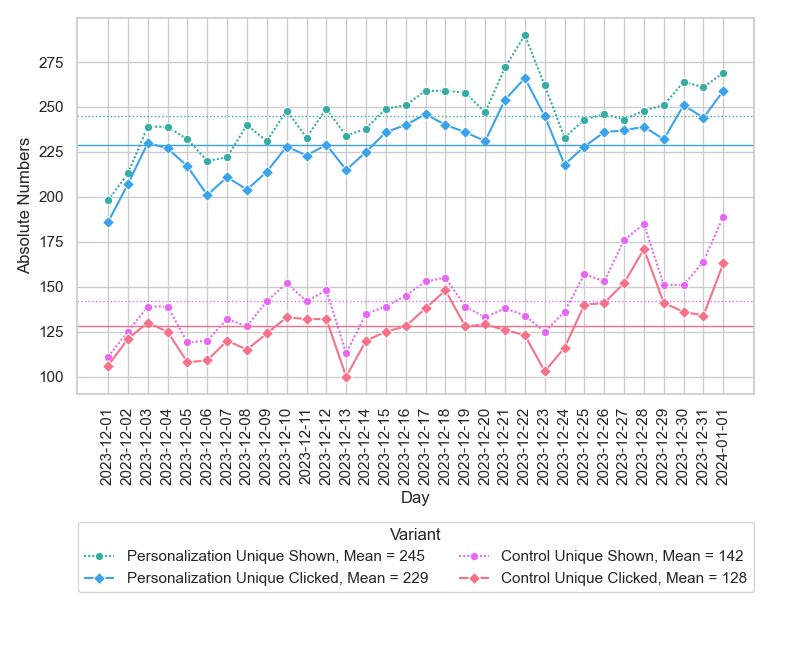}
\caption{Daily click coverage.}
\label{fig:daily_click_coverage_with_mean}
\Description[Daily click coverage per variant]{The daily number of unique clicked articles and the daily number of unique shown articles in the personalization group are consistently higher than in the control group.}
\end{figure}

In Figure~\ref{fig:daily_click_coverage_with_mean}, we plot the number of unique clicked articles---as used for the coverage computation---and the number of unique shown articles over time. We can observe that the number of unique impressed items is consistently higher in the personalization group. On average, in the personalization group, 229 out of 245 uniquely displayed articles per day received at least one click. In contrast, in the control group, 128 out of 142 articles received at least one click. Likewise, the unique number of clicked items is markedly higher as well as in the control group.\footnote{A visualization of the daily click coverage ratio over time can be found in the appendix in Figure~\ref{fig:pdf_click_coverage_lineplot}.} In the long term, this strong positive effect indicates a shift toward better-informed subscribers through personalization.

\paragraph{Popularity}
Popularity bias is commonly considered an undesirable characteristic of recommender systems and it emerges when the system's recommendations are focused too much on already popular items \cite{Klimashevskaia2024Survey}. We therefore analyzed (a) how popular the personalized recommendations are compared to the articles shown in the control group, and (b) if the recommendations have an impact on the average popularity of the articles that readers click on.

Table~\ref{tab:ARP_ACP} shows the results for the three popularity metrics \textbf{average recommendation popularity} (\textbf{ARP}), \textbf{average click popularity} (\textbf{ACP}), and the \textbf{percentage of popular items} (\textbf{PPI}) when considering the data from the entire A/B testing period. For all of them, higher values indicate a higher popularity bias. Our results show that the personalized recommendations actually lead to lower values for all popularity metrics compared to the control group. The recommendations, on average, include less popular items (ARP) and the readers click on these less popular items (ACP and PPI). Considering PPI, we observed a relative decrease of over 4\,\%. A Mann-Whitney U test showed significant differences ($p < 0.001$), although with small effect sizes (Cliff's delta $< - 0.10$; one-sided $95$\,\% bootstrap CI for Cliff's delta upper bound $=-0.093$). We recall that the popularity component of the baseline system puts a comparably strong emphasis on article popularity through the respective weight $w_1$. As such, in our study personalization helps to address popularity bias and feedback loops on the platform to some extent already in the short-term perspective.

\begin{table}[!h]
\centering
\begin{tabular}{lp{2.5cm}p{2.5cm}p{2.5cm}}
 \toprule
 & \textbf{ARP} & \textbf{ACP} & \textbf{PPI} \\
 & \tiny{Average Rec. Pop.} & \tiny{Average Click Popularity} & \tiny{Percentage of Popular Items} \\ \hline
 Control & 421.356 & 222.252 & 0.876\\
 Personalization & 353.471 (-16.11\,\%) & 212.945 (-4.19\,\%) & \textbf{0.840}*** (-4.11\,\%) \\
  \bottomrule
\end{tabular}
\caption{Means of ARP, ACP, and PPI Measured over the Entire A/B Test Period. *** indicates $p < 0.001$.}
\label{tab:ARP_ACP}
\end{table}

Like for the other metrics, we also computed the popularity metrics on a daily basis. The results are presented below in Table~\ref{tab:daily_popularity_metrics}. In terms of the absolute values of ARP and ACP, we notice that the numbers are much lower than when considering the entire testing period (as shown in Table~\ref{tab:ARP_ACP}). This is expected, as articles accumulate clicks over time and the daily measurements include many new articles that appeared on that day. More importantly, however, we observe a strong difference between the personalization and the control group in this analysis, highlighting a strong decrease in popularity bias through the personalized recommendations. In terms of relative numbers, we observed decreases of over 49\,\% for daily ARP, over 44\,\% for daily ACP, and over 4\,\% for daily PPI. Welch's \textit{t}-tests confirmed that the differences for all metrics are significant ($p<0.001$, ARP: two-sided $95$\,\% CI for the difference in population means $\left[-79.567,\;-60.114\right]$, ACP: $\left[-41.712,\;-30.744\right]$, PPI: $\left[-0.049,\;-0.029\right]$), with very strong effect sizes (Cohen's $d$ $>3$), suggesting high practical relevance.

\begin{table}[!h]
\centering
\begin{tabular}{lp{3cm}p{3cm}p{3cm}}
    \toprule
& \textbf{ARP} & \textbf{ACP} & \textbf{PPI} \\
  \hline
 Control & 141.683 & 82.264 & 0.856\\
 Personalization & \textbf{71.843}*** (-49.29\,\%) & \textbf{46.036}*** (-44.04\,\%) & \textbf{0.815}*** (-4.79\,\%) \\
  \bottomrule
\end{tabular}
\caption{Means of ARP, ACP, and PPI Measured Daily. *** indicates $p<0.001$.} 
\label{tab:daily_popularity_metrics}
\end{table}

\begin{figure}[ht]
\centering
\includegraphics[width=0.8\textwidth]{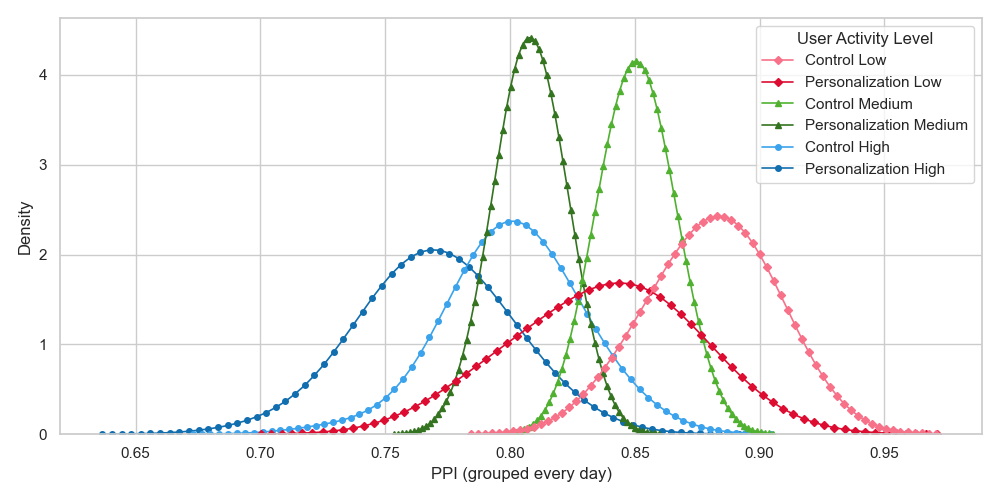}
\caption{Probability density functions daily PPI per user activity level.}
\label{fig:PPI-per-activity-level}
\Description[Probability density functions daily PPI per user activity level]{The probability density functions of daily PPI per user activity level in the personalization group are shifted to the left, indicating lower PPI than in the comparable segments of the control group.}
\end{figure}

We furthermore investigated the effects of personalization on popularity bias for the reader segments with different activity levels. Significant differences (Welch's \textit{t}-test, $p < 0.001$) with strong effect sizes (Cohen's $d$ $<-1$) were found also when performing this subgroup analysis based on the users' activity levels for all metrics. Considering daily PPI per user activity level, we observed a relative decrease of over 4\,\% for low active users, over 5\,\% for medium active users, and almost 4\,\% for high active users. Figure~\ref{fig:PPI-per-activity-level} illustrates the probability density function for the PPI metric, showing clear differences for all activity levels.\footnote{Visualizations for the ARP and ACP metrics can be found in the appendix in Figure~\ref{fig:subfig_ARP} and in Figure~\ref{fig:subfig_ACP}.} The analysis further reveals that popularity bias is highest for the least active user group, and the bias is generally lower the more active users are. This hints to a usage pattern where low-activity users mainly use the app to catch up with the trending news, while more active users also consume content that is more niche.

\paragraph{Summary}
We summarize our findings about the journalistic values measured via diversity, coverage, and popularity metrics, as follows:
\begin{itemize}
\item \textit{Personalization leads to a more diversified exposure of articles and a slightly more diversified consumption behavior}. The impressions and clicks in the personalization group are more evenly distributed across the spectrum of sections, evidenced by a lower Gini index. Linking these findings to \textit{KPI-3} and \textit{KPI-4} on topic diversity in exposure and consumption, we find that personalization can promote a more widespread access to information. Overall, in the long term, a better-informed readership can be expected.

\item \textit{Readers with personalized recommendations explore a broader range of articles per day}. This is evidenced by the increased average click coverage measured daily. Hence, \textit{KPI-5}, which reflects how broadly the subscribers are informed in terms of article coverage, is improved through personalization.

\item \textit{Personalized recommendations, on average, include less popular items (ARP), which is also reflected in the readers' click behavior (ACP, PPI)}. All three popularity metrics show lower values (indicating a lower popularity bias) in the personalization group. Measured daily, a significant decrease in popularity bias through personalized recommendations is observed. Moreover, as users become more active, the popularity bias decreases. Linking these findings to \textit{KPI-6} and \textit{KPI-7}, popularity in exposure and consumption, we find that personalization promotes a better targeted exposure of articles to subscribers and improved visibility of less popular items. Improvements in these KPIs may thus help to successfully avoid feedback loops in the long term.
\end{itemize}

\section{Discussion}\label{sec:discussion}
In this section, we summarize the main insights and implications of our study, discuss research limitations, and review practical challenges of designing and implementing personalized recommendations for legacy media online services.

\subsection{Summary and Implications}
Overall, our analyses show that personalization can have several positive effects even under common constraints of legacy media organizations, which strive to uphold editorial values and journalistic missions by maintaining editorial control over several aspects of the online offering. The analysis of the A/B test outcomes demonstrated that the implementation of personalized recommendations shortens the time for users to identify relevant content, encourages the exploration of a broader range of available content, and results in an overall increase in active reading time. These are factors that can contribute to an improved user experience of the service and to an increased engagement, which may in turn positively impact customer loyalty and continued subscriptions \cite{Li2019Asurvey, Wu2023PersonalizedNRSSurvey, ZHANG2018Therole}. At the same time, after personalization, we found that the consumed content is more diverse and less focused on already popular content, supporting Aftenposten's editorial mission and central journalistic values. These positive effects were observed both for highly active readers and for occasional service users.

Most of our analyses revealed statistically significant improvements that were achieved through personalization, often paired with strong effect sizes. In certain cases, while the results were statistically significant, we  only observed modest or small positive effects of personalization, e.g., in terms of the diversity of the categories of the exposed items. We recall that we do not expect radical changes in reading behavior for different reasons. First, the implemented approach of controlled personalization represents a comparably small intervention regarding the displayed content in general. Second, the duration of the A/B test (34 days) also limits what we can expect in terms of changing reader interests. Therefore, we see the generally positive tendencies as an encouraging outcome that is indicative of future developments in the long term. Generally, even small improvements can be practically relevant for businesses, as they accumulate over time. In this context, sustained and increased exposure to personalized article recommendations is expected to further align content with user preferences, eventually resulting in stronger engagement and more pronounced practical effects.

In sum, our study showed that even adding a small touch of personalization can help to shift the reading behavior of users in desired directions. Thus, we see our analyses as an encouragement for similar legacy media organizations to more often explore opportunities for personalization for their online offering. However, our study also highlights that it is important to track and analyze user behavior in depth and to rely on a set of metrics that capture multiple types of effects of personalization on user behavior. Focusing almost exclusively on click-through-rates, as done frequently in industry and academic research, is probably in most cases insufficient and may be even misleading in some cases \cite{jannachjugovactmis2019}.

\subsection{Research Limitations}
Since our work reports on the outcomes of a particular industry case study, the findings reported in this work reflect the outcomes from one specific experimental configuration. For example, the importance weight of the personalization component was fixed to one particular value that limited the potential impact on the ranking. Furthermore, our study was focused on the subgroup of users of Aftenposten's mobile app. Therefore, further studies are required to assess if our findings generalize for other algorithmic configurations or for users with desktop devices.

In terms of the effects of adding a personalization component to the recommendation system, we found that personalization increased engagement, shifted user behavior in desired directions, and eased the navigation for readers. Commonly, these observations are considered as good predictors for the positive long-term effects, for instance, in terms of customer loyalty. A direct measurement of the effects of personalization on churn rates or the customer lifetime value has not been done so far and needs a much longer observation period due to the delayed effects of recommendations.

The considered A/B testing period lasted for more than one month, which is in the range or even longer of what is reported in previous works that studied news recommender systems `in the wild'. A potential limitation of A/B tests, in particular in the news domain, is that major events such as earthquakes or war breakouts can heavily and instantaneously impact the users' reading behavior. During the time of our A/B test, no such major events occurred. It has to be noted that the last week of the A/B test fell into a time where many users may be on Christmas vacation. We did however not find a notable shift in user behavior, except that the user activity level was generally lower on December 24.

\subsection{Organizational Challenges}\label{subsec:real-world-challenges}
News recommendation faces various well-known \emph{technical} challenges. The dynamic nature of news for example necessitates continuous adaptation due to the high frequency of content updates: articles are regularly added, removed, and revised, demanding frequent model retraining and a robust approach to handling evolving news platforms. Furthermore, the inherent volatility of news, particularly during significant events such as elections, introduces the potential for algorithmic bias and instability \cite{KarimiIPM2018,raza2021news}. Implementing recommender systems within legacy media organizations however presents unique additional challenges that extend beyond mainly technical ones and those encountered in purely digital content platforms such as Netflix or Spotify \cite{Lu2020BeyondOptimizing, Thurman2019My}.

\paragraph{The Role of Editors}
A primary characteristic of legacy media organizations lies in the prioritization of editorial values and journalistic missions \cite{Lu2020BeyondOptimizing, klimashevskaja2024empowering, Bauer2024WhereAre, Gulla2021Recommendingnews}. Unlike platforms optimized solely for user engagement, legacy organizations bear a responsibility to maintain public trust, promote democratic values, and deliver accurate, regularly updated information \cite{Vandenbroucke2024It}. This responsibility translates into a strong desire for editorial control over the recommendation process. Editors actively seek to influence both the selection of articles considered for recommendation and to ensure that content aligns with the organization's journalistic mission. This control is reflected in the careful definition of the candidate set of articles and in the establishment of rules governing algorithmic inclusion. Without these rules, news recommender systems risk becoming too highly tailored and personalized to individual users, resulting in a loss of shared reality among the user base. Moreover, these regulations (and the specific personalized method) not only serve to preserve a shared reality among readers, they mitigate the risk of a divergence in news perception and consumption, avoiding echo chambers.

\paragraph{Competing Objectives}
However, this emphasis on journalistic values often clashes with central business objectives. A key tension exists between short-term performance metrics---such as daily article views---and long-term goals like subscriber growth. This is further complicated by the differing KPIs prioritized by various stakeholders, including journalists, sales departments, and data science teams. The pursuit of profit, often exemplified by paywall strategies, can also undermine the goal of an informed public.\footnote{Let us note here the importance of prioritizing content that benefits society as a whole, rather than solely focusing on maximizing company profit.} For example, while journalists aim for broad readership, the revenue targets of the organization---often achieved through paywalls---can create a barrier to that objective, limiting article accessibility.

Crucially, the set of considered metrics not only have to consider journalistic values, it is also important that regular or periodic data monitoring is implemented to prevent journalists from gaming the system to achieve short-term gains. To ensure consistent evaluation and alignment with these values, a comprehensive and multifaceted analysis of the metrics is required, moving beyond standard engagement metrics to account for journalistic values and societal impact---a point underscored by the findings of this case study. Furthermore, an organization-wide standardization of the defined metrics is necessary to facilitate transparent communication and objective comparison of performance across different stakeholders and over time.

\paragraph{Communication and Transparency}
Finally, organizational barriers can impede successful implementation. Skepticism among journalists regarding algorithmic recommendations, stemming from concerns about filter bubbles and a lack of transparency, requires careful communication and collaboration between editorial, product, data, and curation teams. Building trust and fostering a shared understanding of the system's functionality and limitations is essential to overcoming these perceptual challenges. Gradual and iterative improvement based on collective decisions, as well as the alignment of efforts toward a shared and transparently communicated objective, are key aspects to successfully implement news recommender systems.

\section{Conclusion and Outlook}\label{sec:conclusion}
Personalized news recommender systems have been extensively explored for major news aggregator sites such as Google News or Yahoo! News. With our study, we showcased that personalization can also be effective in the particular environment of legacy media online services. Importantly, however, we emphasize that multi-dimensional analyses are needed to fully understand the effects of personalization on users. We hope that our study can serve as a blueprint for multi-dimensional analyses also in similar legacy news organizations.

We see various directions for expanding our research in the future. From a technical perspective, these future works include the exploration of alternative recommendation strategies and adding more weights to the personalization component. In terms of future analyses, depending on data availability, we plan to conduct additional subgroup analyses, taking in particular into account factors such as user demographics. Finally, from a business perspective, we aim at conducting additional A/B tests over longer periods of time and study business-related aspects such as customer retention and churn behavior from a longitudinal perspective.

\section*{Data Availability Statement}
Due to company-internal data protection regulations, we cannot share the data that was used for the analyses described in the article.

\section*{Acknowledgment}
This research was supported by industry partners and the Research Council of Norway with funding to MediaFutures: Research Centre for Responsible Media Technology and Innovation, through the Centres for Research-based Innovation scheme, project number 309339. Furthermore, this research was funded in part by the Austrian Science Fund (FWF) 10.55776/COE12. For open access purposes, the author has applied a CC BY public copyright license to any author accepted manuscript version arising from this submission.

\bibliographystyle{ACM-Reference-Format}

\newpage
\section*{Appendix}

\begingroup
In the following, we present the distribution of activity durations per ranking strategy in Figure~\ref{fig:ad_control} and in Figure~\ref{fig:ad_personalization}. As illustrated in Figure~\ref{fig:ad}, the distribution of activity durations per user activity level is presented. In Figures~\ref{fig:subfig_ARP} and~\ref{fig:subfig_ACP}, the visualizations for the ARP and ACP metrics per user activity level are shown, indicating clear differences for all activity levels. To provide a clearer understanding of the daily click coverage metric, we present the probability density functions for both groups in Figure~\ref{fig:pdf_click_coverage_kde} and depict the longitudinal trajectory of this metric in Figure~\ref{fig:pdf_click_coverage_lineplot}.

\vspace{-2mm}

\begin{figure}[H]
\centering
\begin{minipage}{0.49\textwidth}
\centering
\includegraphics[width=0.98\linewidth]{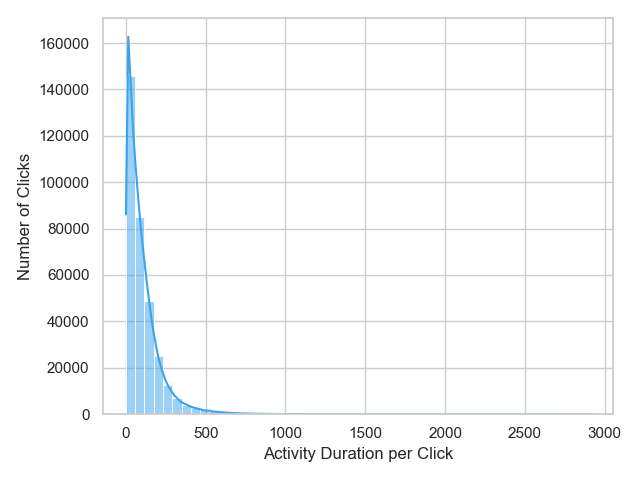}
\caption{Distribution of activity durations for the control group.}
\label{fig:ad_control}
\Description[Distribution of activity durations for the control group]{The distribution of activity durations for the control group shows that a large number of users only briefly interact with the articles they click.}
\end{minipage}
\hfill
\begin{minipage}{0.49\textwidth}
\centering
\includegraphics[width=0.98\linewidth]{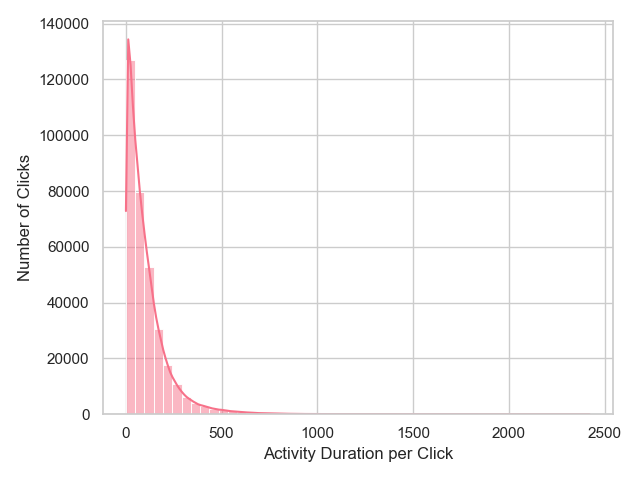}
\caption{Distribution of activity durations for the personalization group.}
\label{fig:ad_personalization}
\Description[Distribution of activity durations for the personalization group]{The distribution of activity durations for the personalization group shows that a large number of users only briefly interact with the articles they click.}
\end{minipage}

\vspace{8mm}

\centering
\includegraphics[width=0.98\textwidth]{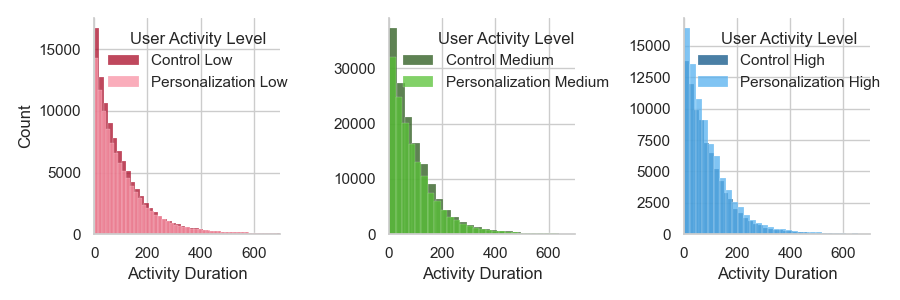} \\
\caption{Distribution of activity durations per user activity level and ranking strategy.}
\label{fig:ad}
\Description[Distribution of activity durations per user activity level and ranking strategy]{The distribution of activity durations for low activity users shows that readers in the personalization group interact with clicked articles slightly longer than those in the control group. The same holds for high activity users, but not for the medium group.}
\end{figure}

\newpage

\begin{figure}[H]
\centering
\includegraphics[width=0.8\textwidth]{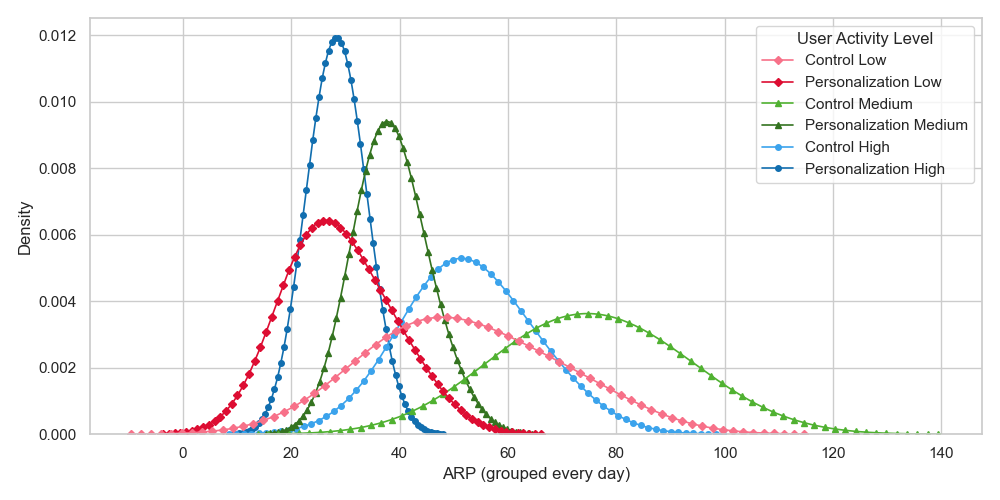}
\caption{Probability density functions daily ARP per user activity level.}
\label{fig:subfig_ARP}
\Description[Probability density functions daily ARP per user activity level]{The probability density functions of daily ARP per user activity level in the personalization group are shifted to the left, indicating lower ARP than in the comparable segments of the control group.}
\end{figure}

\vspace{-10pt}

\begin{figure}[H]
\vspace*{2pt}
\centering
\includegraphics[width=0.8\textwidth]{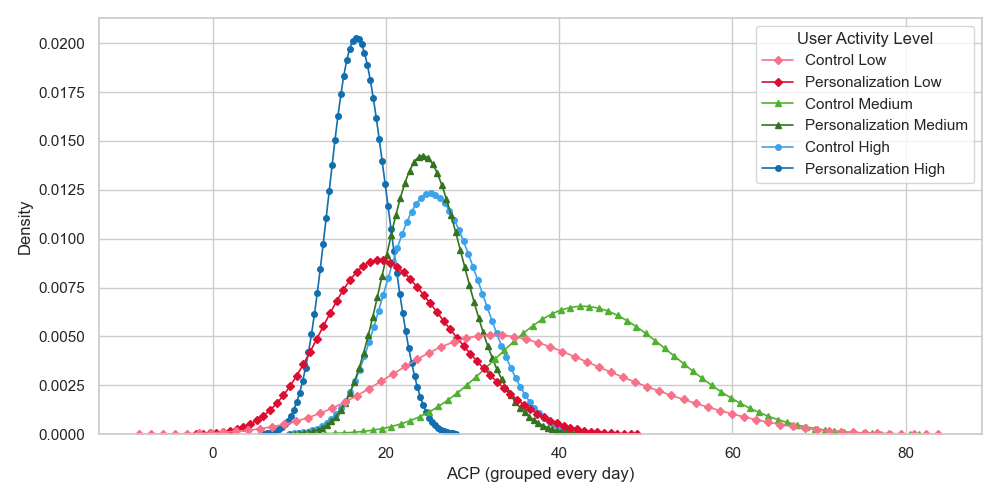}
\caption{Probability density functions daily ACP per user activity level.}
\label{fig:subfig_ACP}
\Description[Probability density functions daily ACP per user activity level]{The probability density functions of daily ACP per user activity level in the personalization group are shifted to the left, indicating lower ACP than in the comparable segments of the control group.}
\end{figure}

\begin{figure}[H]
\centering
\includegraphics[height=5cm,keepaspectratio]{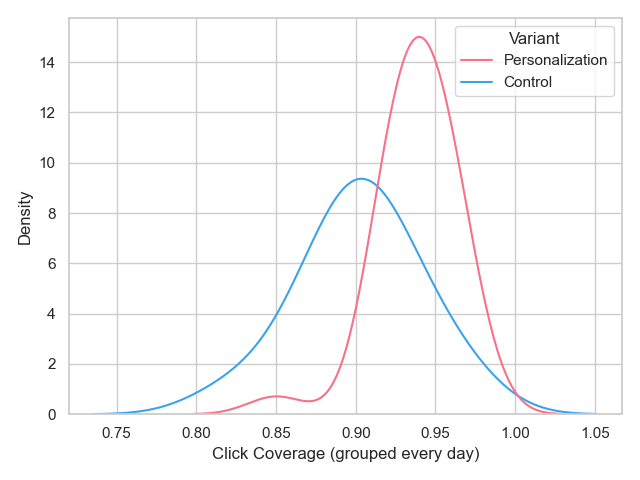}
\caption{Probability density function of daily click coverage.}
\label{fig:pdf_click_coverage_kde}
\Description[Probability density function of daily click coverage]{The probability density function of daily click coverage in the personalization group is shifted to the right, indicating higher click coverage than in the control group.}
\end{figure}

\begin{figure}[H]
\centering
\includegraphics[width=0.8\textwidth,keepaspectratio]{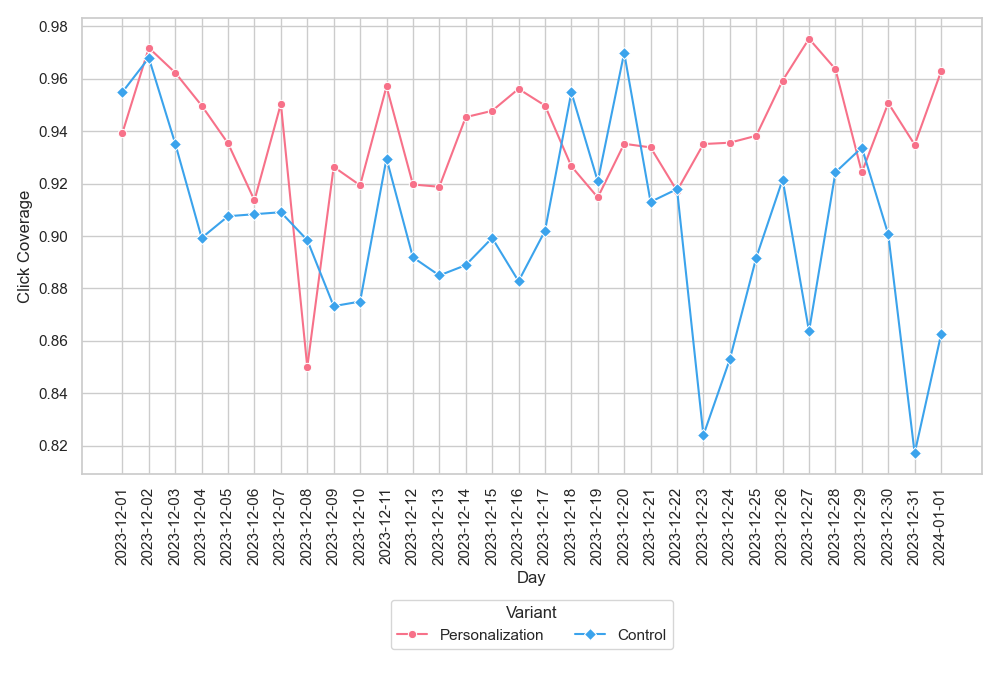}
\caption{Daily click coverage.}
\label{fig:pdf_click_coverage_lineplot}
\Description[Daily click coverage]{The lineplot of daily click coverage per variant shows almost constantly higher values in the personalization group, indicating higher click coverage than in the control group.}
\end{figure}

\endgroup

\end{document}